\documentclass{aa} %\documentclass[structabstract]{aa}

\usepackage{graphicx}
\usepackage{lscape}
\usepackage{txfonts}
\usepackage{times}
\usepackage{color}

\usepackage{natbib}
\bibpunct{(}{)}{;}{a}{}{,} % to follow the A&A style, natbib cite format used by A&A and ApJ

\def\V0{$V_0$}
\def\V{$V$}
\def\Eby{$E(b-y)$ }
\def\EBV{$E(B-V)$ }
\def\Mv{$M_V$ }
\def\Av{$A_V$ }
\def\b-y{$b-y$ }
\def\c1{$c_1$ }
\def\m1{$m_1$ }
\def\BrM1{$[m_{1}]$ }
\def\BrC1{$[c_{1}]$ }
\def\'{$ ^{\rm '}$ }
\def\C0{$c_0$ }
\def\M0{$m_0$ }
\def\by0{$(b-y)_0$ }
\def\uvbyb{$uvby\beta$ }
\def\BV0{$(B-V)_0$ }

\newcommand{\hi}{H\,{\sc i}}
\newcommand{\hii}{H\,{\sc ii}}
\newcommand{\Ha}{H$\alpha$}

\begin{document}

 \title{Massive Stellar Content of the Galactic Supershell GSH~305+01-24
       \thanks{Tables 1 and 3 are only available in electronic form at the CDS via anonymous ftp 
			         to cdsarc.u-strasbg.fr (130.79.128.5) or via http://cdsweb.u-strasbg.fr/cgi-bin/qcat?J/A+A/.../
							}
        }

 \author{N. T. Kaltcheva\inst{1}
               \and
               V. K. Golev\inst{2}
               \and
               K. Moran\inst{1}
         }

   \institute{Department of Physics and Astronomy, University of Wisconsin Oshkosh, 800 Algoma Blvd.,
	            Oshkosh, WI 54901, USA \\
              \email{kaltchev@uwosh.edu}
              \and
              Department of Astronomy, Faculty of Physics, St Kliment Ohridski University of Sofia,
							5 James Bourchier Blvd., BG-1164 Sofia, Bulgaria \\
              \email{valgol@phys.uni-sofia.bg}
             }

  \date{Received 13 March 2013 / Accepted 02 December 2013}
  \offprints{kaltchev@uwosh.edu}
  \titlerunning{Centaurus Star-Forming Field}
  \authorrunning{Kaltcheva et al.}
	
  % \abstract{}{}{}{}{}
  % 5 {} token are mandatory

  \abstract
  % context heading (optional), leave it empty if necessary
	{}
  {The distribution of OB stars along with that of \Ha, $^{12}$CO, dust
   infrared emission, and neutral hydrogen is carried out in order to provide
   a more complete picture of interactions of the young massive stars and the observed supershell GSH 305+01$-$24.
  }
  {The studied field is located between $299^\circ \le l \le 311^\circ$ and $-5^\circ \le b \le 7^\circ$.
	 The investigation is based on nearly 700 O-B9 stars with \uvbyb photometry currently available.
   The derived stellar physical parameters were used to establish a   homogeneous scale for the distances
	 and extinction of light for major apparent groups and layers of foreground and background stars in
	 Centaurus and study the interaction with the surrounding interstellar medium.
  }
  {The distance to the entire Centaurus star-forming complex is revised and a  maximum of the OB-star
	 distance distribution is found at 1.8$\pm$0.4 (r.m.s) kpc.
	 The massive star component of GSH 305+01$-$24 is identified at about 85-90 \% completeness up to 11.5-12 mag.
	 The projected coincidence of the OB stars with the shell and the similarities between the shell's morphology
	 and the OB-star distribution  indicate a strong interaction of the stellar winds with the superbubble material.
	 We demonstrate that these stars contribute a sufficient wind injection energy in order to explain the observed
	 size and expansion velocity of the supershell.
	 The derived stellar ages suggest an age gradient over the Coalsack Loop.
	 A continuous star-formation might be taking place within the shell with the youngest stars located at
	 its periphery and the open cluster NGC 4755 being the oldest.
	 A layer of very young stars at 1 kpc is detected and its connection to both GSH 305+01$-$24 and
	 the foreground GSH 304$-$00$-$12 \hi\ shells is investigated.
	}
	{}

  \maketitle

  %
  %________________________________________________________________
	
  \section{Introduction}\label{sec:intro}

  The Cen OB1 association represents one of the prominent star-forming regions of the Milky Way.
	It forms part of an extended star-formation complex in the disk between Galactic longitudes $299^\circ-311^\circ$.
	The field is dominated by the \hi\ supershell GSH 305+01$-$24  \citep{mcc01} and its optical counterpart,
	the Coalsack Loop, first reported by \citet{wal98}.
	In both two-armed and four-armed models of the Galaxy the field is associated with the Carina-Sagittarius arm
	or with an inner spur of the arm (see for example \citealt{hum74}).
	The field is located in the direction of the Lower Centaurus-Crux (LCC) association, which is one of the
	most prominent tracers of the Gould Belt and is likely to be our closest giant star-forming complex
	(Comer$\acute{\mathrm o}$n \citeyear{com01}), that could also represent a large-scale complex in the
	region of the Orion Arm \citep{sar03, deZ99, eli09}.
	
	% Fig 1 *******************************************************
  %
  \begin{figure*}
  \centering
  \includegraphics[width=45pc]{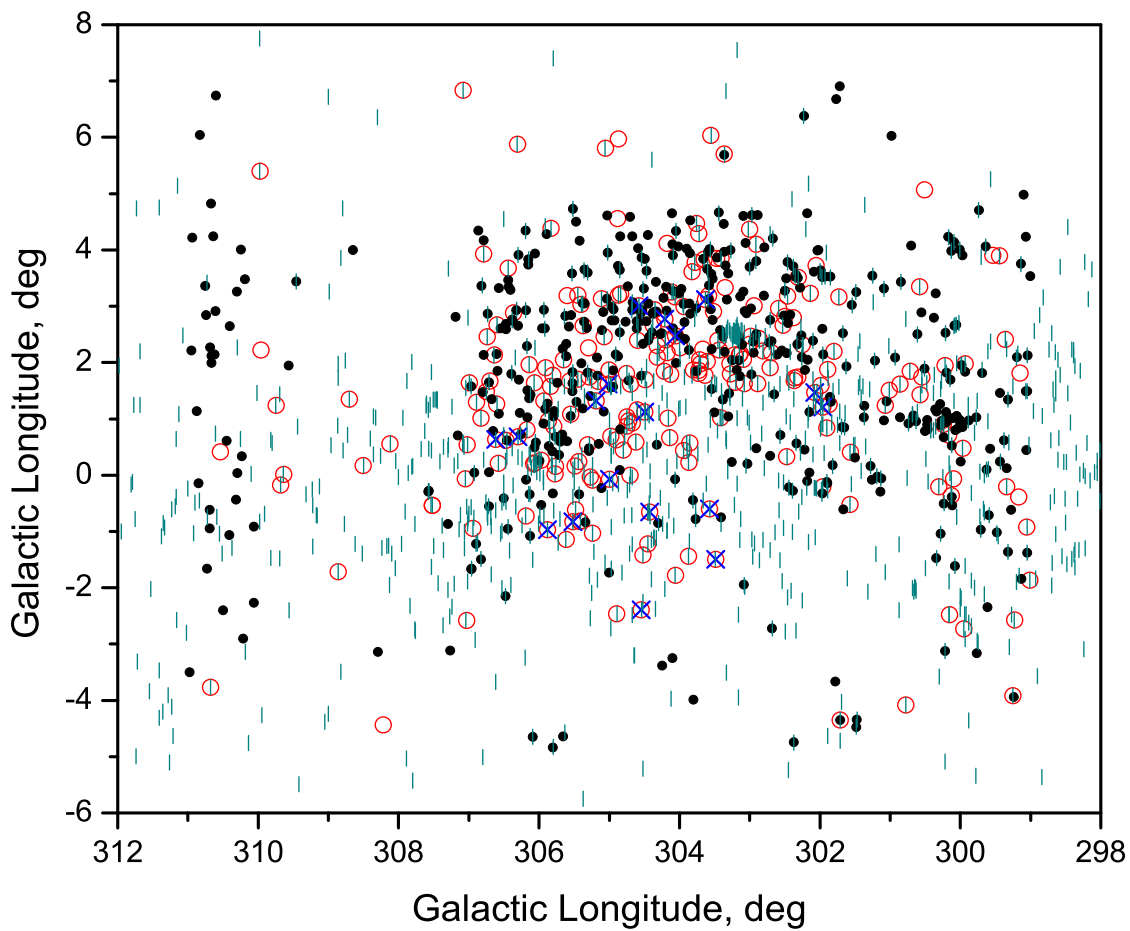}
  \caption[]{The sample of 693 stars plotted in Galactic coordinates.
	  Intrinsically bright stars with $M_V < -2$ mag
		(which are also located farther than 1 kpc)
		are marked by red open circles and stars intrinsically fainter than $-2$ mag (which are all
		found closer than 1 kpc) by black filled circles.
		Stars from the classical Cen OB1 association are marked by blue x-symbols.
		The stars from the {\it Luminous Stars} catalog (\citealt{ree03}) are represented by green vertical symbols.
    The concentrations represent NGC 4755 ($l,b = 303^{\circ}\!\!.206, 2^{\circ}\!\!.59$) and
	  St 16 ($l,b = 306^{\circ}\!\!.15, 0^{\circ}\!\!.0656$).
		See details in paragraph~\ref{sec:distOB}.
		See the electronic edition for a color version of the figure.
    }
 \label{fig1}
 \end{figure*}
	
 The direction between Galactic longitudes $302^\circ$ and $313^\circ$ is tangent to the Scutum-Crux arm
 (also referred to as the Scutum-Centaurus, or Centaurus arm), corresponding to maxima in the
 thermal radio continuum, \hi\ and CO emissions \citep{tay93, blo90, bro92, ben08}.
 As one of the major Galactic star-forming regions, Centaurus can be used to model and/or test various
 spiral morphologies: from grand-design logarithmic spirals (\citealt{rus03}; Vall$\acute{\mathrm e}$e
 \citeyear{val08}) to designs dominated by rings or pseudo-rings (see \citealt{rau10} for a recent discussion).
 The Centaurus field may be particularly important for understanding a spiral design with two principal arms
 and underlying enhancements in the old stellar disk, recently discussed by L$\acute{\mathrm e}$pine et al. 
 (\citeyear{lep01}) and \citet{ben08}.

 There have been several attempts to establish the distance to Cen OB1 \citep{hum78,bla89,gar92,mel95},
 most based on the photometric data of \citet{bla89}, or earlier studies.
 All estimates use {\it UBV} photometry and spectro\-photometric calibrations, and place the association
 1.9 to 2.5 kpc from the Sun.
 Recently, using proper motion and spectrophotometric distances, \citet{cor12} compiled a list of 56 members of
 the association and found an average distance of $2.6\pm0.4$ kpc.

 In the present study we combine surveys of \Ha, $^{12}$CO, dust infrared emission at 100 $\mu$m, and
 neutral hydrogen with intermediate-band \uvbyb\ photometry to examine the correlation between the location
 of the OB-stars and the neutral and ionized material in the GSH 305+01$-$24 supershell.
 Based on homogeneous distances of nearly 700 early-type stars, we select spatially coherent stellar groupings
 and revise the classical concept of the Cen OB1 association.

 This study is organized as follows.
 In Sections \ref{sec:sample} and \ref{sec:distOB} we discuss the stellar sample and findings regarding
 groupings and the overall distribution of the O-B9 stars toward Centaurus.
 Section \ref{sec:correlation} is focused on the census of OB stars within GSH 305+01$-$24 and
 on the correlation between massive stars and interstellar material.
 Brief concluding remarks are presented in Section \ref{sec:concl}.
	
 %
 %________________________________________________________________
	
 \section{The stellar sample}\label{sec:sample}
	
 Among the wide variety of photometric systems available today the \uvbyb\ photometry is arguably better
 suited for the study of individual stars in terms of stellar luminosities and distance moduli than any
 other photometric system in wide use.
 A detailed discussion of the \uvbyb\ system can be found in \citet{str66}.
 In the O-B9 spectral range considered in this study the H$\beta$ index is a measure for the strength of
 the H$\beta$ line, and thus a luminosity indicator that is primary used to calculate \Mv.
 The $(b-y)$ index is a temperature indicator, providing the true stellar color and thus the interstellar reddening.
 The color difference $c_1= (u-v)-(v-b)$ is a measure of the Balmer discontinuity and an indicator
 for the effective temperature in O-B9 stars, which is also used in the \Mv calculation.
 The color difference $m_1= (v-b)-(b-y)$  measures the blocking by metallic lines, but it is not used
 in any of the calibrations applicable to O-B9 stars.
 A recent description of the \uvbyb\ photometric quantities is provided
 by $\acute{\mathrm A}$rnad$\acute{\mathrm o}$ttir et al. (\citeyear{arn10}).
	
 The sample for the present study contains 693 (mostly) field stars within the coordinate range
 $299^\circ \le l \le 311^\circ$ and $-5^\circ \le b \le  +7^\circ$.
 The stars are of spectral type O to B9 and have complete \uvbyb\ photometry.
 Some of the brightest stars of several loose clusters, such as St~16, are included in the sample,
 as listed in Table~1, available only electronically.
 In addition, 56 members of the young open cluster NGC~4755 ($\kappa$ Crucis cluster, Jewel Box) are considered.
 All \uvbyb\ photometry was extracted from the catalog of \citet{hau98}.
 The homogeneity of similar samples is discussed in earlier papers that studied the structure of Galactic
 star-forming regions \citep[see for example][]{kal12}.

 %
 %________________________________________________________________	
 \subsection{Completeness of the Stellar Sample}

 The locations of the 693 sample stars in Galactic coordinates are shown in Fig.~\ref{fig1}.
 When selecting the stars we considered a field larger than the projected size of GSH 305+01$-$24.
 This was done with purpose to select all massive stars projected towards the supershell and its environments,
 i.e. the entire Centaurus field.  Fig.~\ref{fig1} shows the 693 sample stars which are marked with two
 different symbols, according to the separation we establish in Section \ref{sec:distOB}.
 There we separate the sample into closer than 1 kpc and intrinsically fainter than $M_V$ of $-2$ mag stars
 (black filled symbols) and distant, intrinsically brighter than $-2$ mag stars (open circles).
 The stars of the classical Cen OB1 association are overplotted with x-symbols.

 The best possible way to estimate the sample completeness is to compare to the {\it Luminous Stars} (LS) Catalog
 (see \citealp{ste71,ree03}).
 In the coordinate range $l = (299^\circ,311^\circ), b = (-6^\circ, 8^\circ)$, 962 LS are listed,
 and are marked with green vertical symbols in Fig.~\ref{fig1}.
 Of these, 172 correspond to stars intrinsically fainter, and 180 to stars intrinsically brighter according
 to our separation criterion (see Section\ref{sec:distOB}).
 This implies that the LS catalog contains stars that are intrinsically faint and not connected to the star-forming field.

 In order to estimate the completeness of the stellar sample projected toward the GSH 305+01$-$24 supershell
 we restrict the coordinate range to $l = (300^\circ,307^\circ), b = (-5^\circ, 7^\circ)$.
 If NGC~4755 is excluded, there are 675 LS stars in this region.
 Of them 152 correspond to sample stars intrinsically fainter than $M_V = -2$ mag (393 total),
 and 158 to intrinsically brighter stars (202 in total).
 It seems reasonable to assume that the same proportionality between number of stars  brighter and fainter
 than $-2$ mag is true for all 675 LS stars projected toward the shell.
 That would imply that the sample located beyond 1 kpc is at least $70-80$\% complete to the limit of the LS catalog.
 In addition, our sample  contains nearly 50 intrinsically bright distant stars not identified as LS stars,
 which should improve its completeness.
  		
 One may note that there are not many \uvbyb\ data toward the longitude interval $307^\circ-310^\circ$ in Fig.~\ref{fig1}.
 This might indicate incompleteness due to the fact that the \citet{hau98} catalog is a compilation from different
 sources with different completeness levels depending on the position on the sky.
 On the other hand, the LS catalog which is complete to $m_{ph} \sim 12$ limit shows to some extent a similar behavior,
 suggesting absence of prominent star formation in this direction.
 In any case, this region is outside the cavity of the \hi\ and \hii\ shells, so it should not affect our findings below.

 %
 %________________________________________________________________
	
 \subsection{The interstellar extinction and distances}
	
 % Table 2 *******************************************************
	
 \setcounter{table}{1}
 \begin{table*}
 \centering
 \caption{Color excess and distance modulus as a function of luminosity class
    (LC) obtained via the $M_V$ calibrations of  \citet{zha83} and \citet{bal84} for HD 114122 as an example.}
 \tiny
 \vspace{0.1in}
 \begin{tabular}{lllllllllll}

  LC 	   & $\beta$ & $(b-y)_0$ & $c_0$ 	  & \Eby\  & 	$V_0$ 	& $M_V$ (Zhang) & DM (Zhang) & 	$M_V$ (B\&S)	& DM (B\&S) \\
	       & 		     &           & 		      &        &  		  	&               & 	      	 & 	            	&        	\\
  V-III  & 2.564 	 & $-0.127$	 & $-0.107$ & 0.641  & 	5.824 	& $-5.414$	    & 11.238     & $-5.41$	      & 11.24 	\\
  II 	   & 2.564 	 & $-0.127$  & $-0.107$ & 0.641  & 	5.824 	& $-5.845$	    & 11.669 	   & $-5.41$	      & 11.24 	\\
  Ib 	   & 2.564   & $-0.096$  & $-0.101$ & 0.61   & 	5.958 	& $-5.862$	    & 11.82 	   & $-5.38$	      & 11.34 	\\
  Iab 	 & 2.564 	 & $-0.069$	 & $-0.096$ & 0.583  & 	6.073 	& $-5.875$	    & 11.948 	   & $-5.36$	      & 11.43 	\\
  Ia 	   & 2.564 	 & $-0.053$	 & $-0.093$ & 0.567  & 	6.143 	& $-5.884$	    & 12.026 	   & $-5.34$	      & 11.48 	\\
	
 \end{tabular}
 \label{tab2}
 \end{table*}	
	
 % Fig 2 *******************************************************
 %
 \begin{figure*}
 \centering
 \includegraphics[width=14pc]{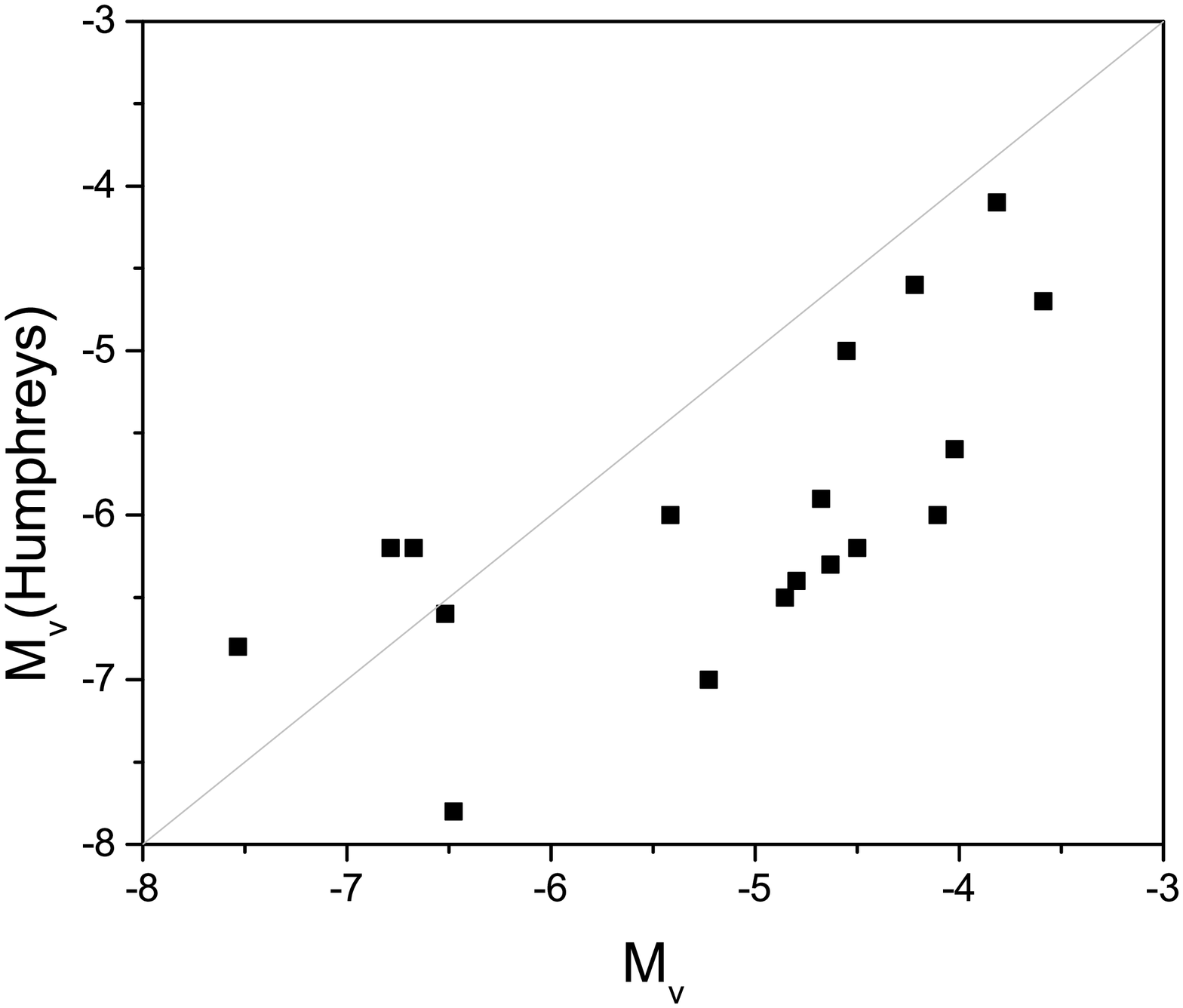}
 \includegraphics[width=14pc]{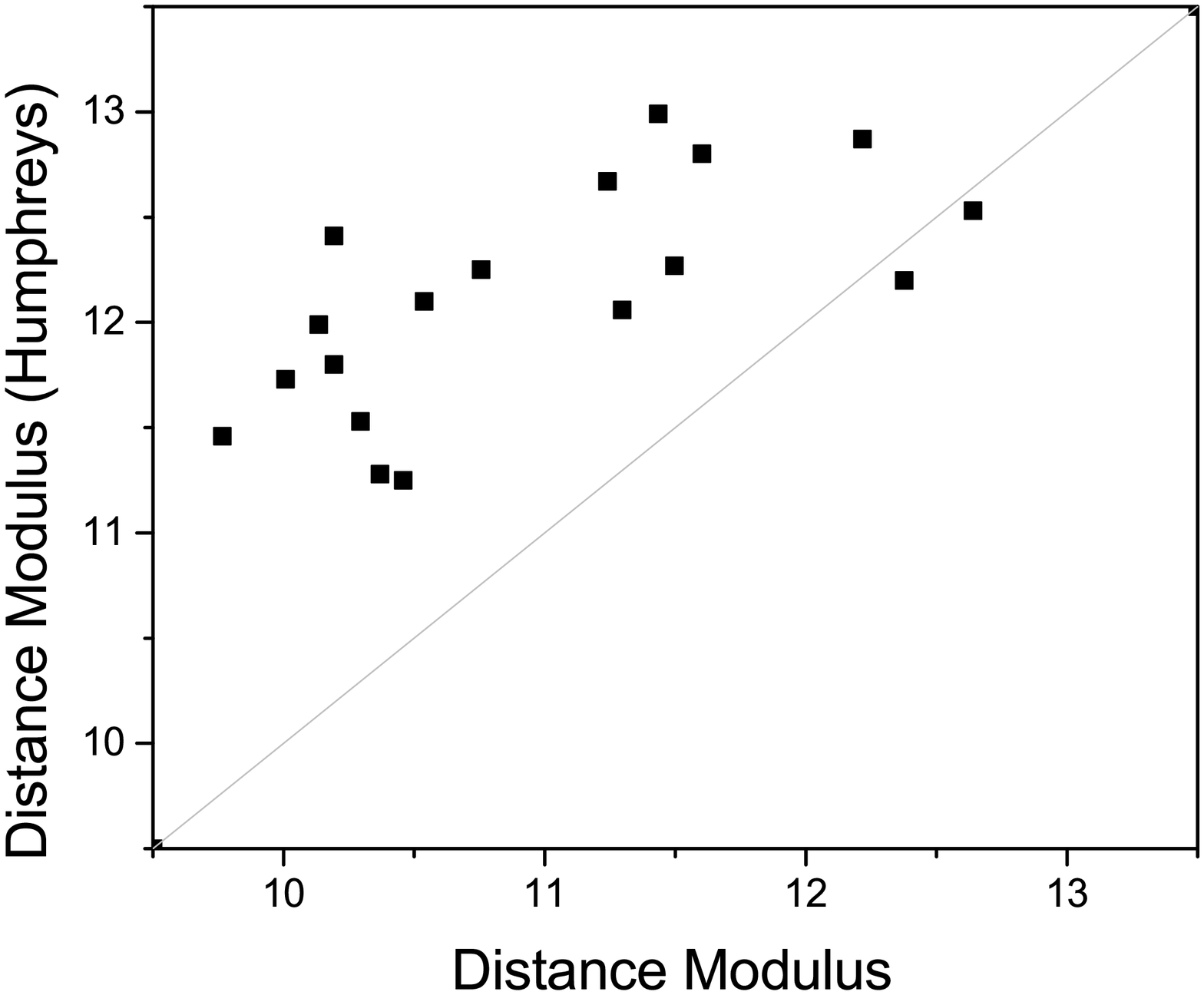}
 \includegraphics[width=14pc]{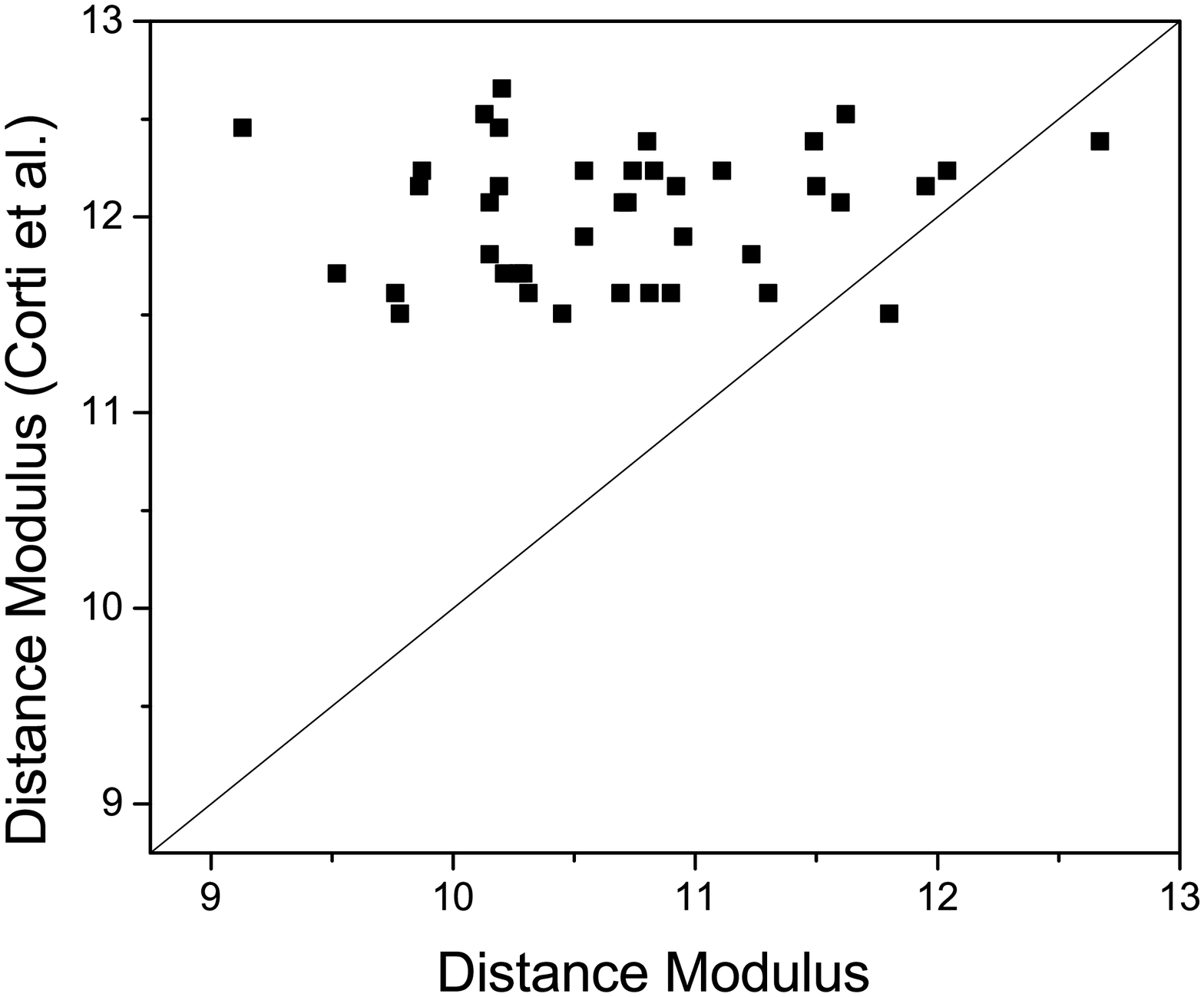}
 \caption[]{Comparison of \uvbyb\ absolute magnitudes and distance moduli obtained here (plotted on the abscissa)
   with previously published  spectroscopic absolute magnitudes and distance moduli.
 }
 \label{fig2}
 \end{figure*}	
	
 The procedure used to obtain color excesses and distances for sample stars is described in detail in \citet{kal00}.
 The color excesses for Luminosity Class (hereafter LC) III, IV, and V were obtained via
 Crawford's \citeyear{cra78} calibration.
 The calibration by \citet{kiw85} was used for LC II, Ib, Iab, and Ia.
 The relations ${\mathrm R}=3.2$ and \EBV\ = \Eby/$\,0.74$ were adopted to obtain $V_0$.
 The calibration by \citet{bal84} was utilized for all O-B9 stars to derive values of $M_V$.
 Since the stars are of early spectral type, the presence of emission lines in their spectra is the
 largest source of error in the calculated absolute magnitudes.
 However, the $\beta$ vs. \C0 diagram (not shown here) indicates that only 7 stars in the sample
 deviate from the main sequence or have photometry affected by emission.
 For all stars with observed $\beta$ lying outside the limits of the $M_V$($\beta$,$c_0$) calibration,
 and for all known emission-line stars, $\beta$ was calculated from \C0
 (see for details \citealt{bal94} and \citealt{kal00}).
 Note that such a procedure yields distance in excellent agreement with the recalculated $Hipparcos$ data
 (see for example \citealt{kal09} and \citealt{kal07}).
		
 The photometric data and the derived stellar parameters are summarized in Table~1 followed by the
 Galactic coordinates, MK type, \uvbyb photometric data, color excess, dereddened photometry,
 calculated absolute magnitude, and true distance modulus.
 The uncertainties in \Mv are of the order of $\pm\,0.3$ mag for O and B types of LC III-V and
 $\pm\,0.5$ mag for B-type supergiants \citep{bal84}.
 An uncertainty of $\pm\,0.3$ mag in $M_V$ propagates to an asymmetric error of $-13$\% to +15\%, and
 uncertainties of $\pm\,0.5$ mag result in $-21$\% to +26\% error in the derived distances.

 Since the photometry used in this paper comes from different sources,
 the homogeneity of the sample is an important issue.
 The above mentioned errors include the possible systematic errors in the existing photometric data.
 Comparisons of existing \uvbyb\ datasets collected by various authors are in general
 in a good agreement \citep[see][]{koc00}.
 The estimated uncertainty in the calculated stellar distances due to possible systematic deviations could
 not exceed $3-5\,$\%.
 Similar approach has been followed by \citet{tor00} who also use the \uvbyb\ photometry from \citet{hau98}
 to derive individual distances and age in their study of the local kinematics of young stars.
 Depending on the spectral type and luminosity class, the errors estimated by \citet{tor00} range
 between 14\% and 23\%.
 It should be noted that in the \uvbyb system both the color excess and absolute magnitude
 calculations do not rely on a precise determination of spectral sub-type, since the calculations are
 carried out in the same manner for types from O to B9, depending on LC only for the color excess derivation.

 In order to infer the physical stellar parameters from the photometry, the spectral and luminosity classifications
 were extracted from the SIMBAD database and checked against the classification from the \BrC1 vs. \BrM1
 diagram (not shown here). 
 Only 3 stars showing disagreement between the photometric quantification and SIMBAD MKK types
 were removed from the sample.
 Great care was taken to resolve all cases of suspected luminosity class misclassification, because
 different calibrations are used to calculate the color excesses for different LC types.
 For stars of LC types I and II we compared the LC accepted in SIMBAD to all of the literature sources of
 luminosity classification as well as to the photometric quantification and we did not find disagreements.

 The Cen~OB1 association was previously studied in the \uvbyb system by \citet{kal94} using all 28 early-type
 members identified by \citet{hum78}.
 The present work offers several improvements to that study.
 Many more stars are included here utilizing homogenized \uvbyb photometry from the \citet{hau98} catalog.
 New MKK classifications made available in the last decade permit a refinement of the LC classification for
 a number of stars.
 Most importantly, the \Mv calibration of \citet{bal84} is applied to all stars, instead of the \citet{zha83}
 calibration used by \citet{kal94} for stars of LC II and I.
 The \citet{bal84} calibration has been tested extensively using $Hipparcos$ data and is applicable to
 evolved stars as well as dwarfs.
 Our extensive tests of both calibrations show that the relations of \citet{zha83} overestimate brightness
 for LC I and II stars.
 Since the majority of members identified by \citet{hum78} are giants and supergiants, such a change
 significantly affects the average distance to the association.

 Inclusion of all photometric and spectral classification data available at present helps to resolve cases of
 stars with controversial LC.
 For example, HD 114122 has four spectral classifications in SIMBAD: O+, B1Iab, B0III, and
 B0.5Ia-ab (the latter adopted by \citealp{hum78} and by \citealp{kal94}).
 An inspection of the star's location in the $[c_1]/[m_1]$ and $\beta/[u-b]$ and $\beta/[c_1]$ diagrams
 (not shown here) indicates LC II-III.
 Table \ref{tab2} presents a summary of color excess E(b-y) and \Mv calculations for the star for
 different LCs with the \Mv calibrations noted above (\citealp{zha83} and \citealp{bal84}).
 It can be noted that the DM differs by only 0.2 mag between LC V-III and LC Ia for the \citet{bal84}
 calibration (for the O-B1 spectral range \Mv is almost the same for all LC).
 The calibration of \citet{zha83} provides intrinsically brighter \Mv in the case of LC II and I.
 In her study, \citet{hum78} lists \Mv$ = -6$ and \noindent DM$\ = 12.67$ (\Av$ = 2.61$).
 Our calculations yield a slightly larger \Mv (regardless of \Mv calibration used) and a smaller DM,
 respectively (DM$\ = 11.24$ mag was adopted here).
 A similar analysis was performed for all stars in the sample with controversial LC designations.

 Careful consideration of LC classification for sample stars and use of a more appropriate \Mv calibration
 result in better distance determinations, thereby helping to
 distinguish between closely spaced groups lying along the line of sight.

 %
 %________________________________________________________________

 \section{Distribution of O-B9 stars toward Centaurus}\label{sec:distOB}

 \citet{hum78} lists 29 stars as members of Cen~OB1 at $l,b = (303.5,2.4)$\ with a distance modulus of
 12 mag (2512 pc).
 In their new delineation of Galactic OB associations \citet{mel95} propose five groups (Cen 1A to Cen 1E)
 located between $303^\circ \le l \le 306^\circ$ at distances varying from 1.8 to 2.2 kpc.
 Their Cen 1A group consists of 37 stars, almost all belonging to NGC~4755.
 That group includes the O9.5-type star HD 311999 seen toward the \hii\ region Gum 46.
 The groups Cen 1B, Cen 1C, and Cen 1D consist of several field stars, while Cen 1E (6 stars) contains
 mostly members of the open cluster St 16.
 Recently, based on proper motion and spectrophotometric distances, \citet{cor12} provide an updated list
 of 56 members of Cen OB1 at an average distance of 2.6{$\pm0.4$} kpc.
 Fig.~\ref{fig2} presents a comparison of the \uvbyb distance moduli obtained here to previously published
 spectroscopic estimates (\citealp{hum78,cor12}).
 The spectroscopic distance moduli appear to be larger as a result of overestimated intrinsic brightness in
 some of the existing spectroscopic calibrations (see for example \citealt{kal11}).
 The distance moduli for members of Cen OB1 selected by \citet{cor12} are all grouped around 12 mag as a
 consequence of the selection procedure adopted by the authors, which considered only stars with consistent
 proper motions located around the DM of 12 mag adopted by \citet{hum78}.
	
 In the present study a homogeneous \uvbyb photometric distance scale was established for nearly 700 O-B9
 field stars and the rich open cluster NGC 4755, permitting us to identify spatially coherent groups
 and layers and to study their characteristics.
				
 % Fig 3 *******************************************************
 %
 \begin{figure*}
 \centering
 \includegraphics[width=18pc]{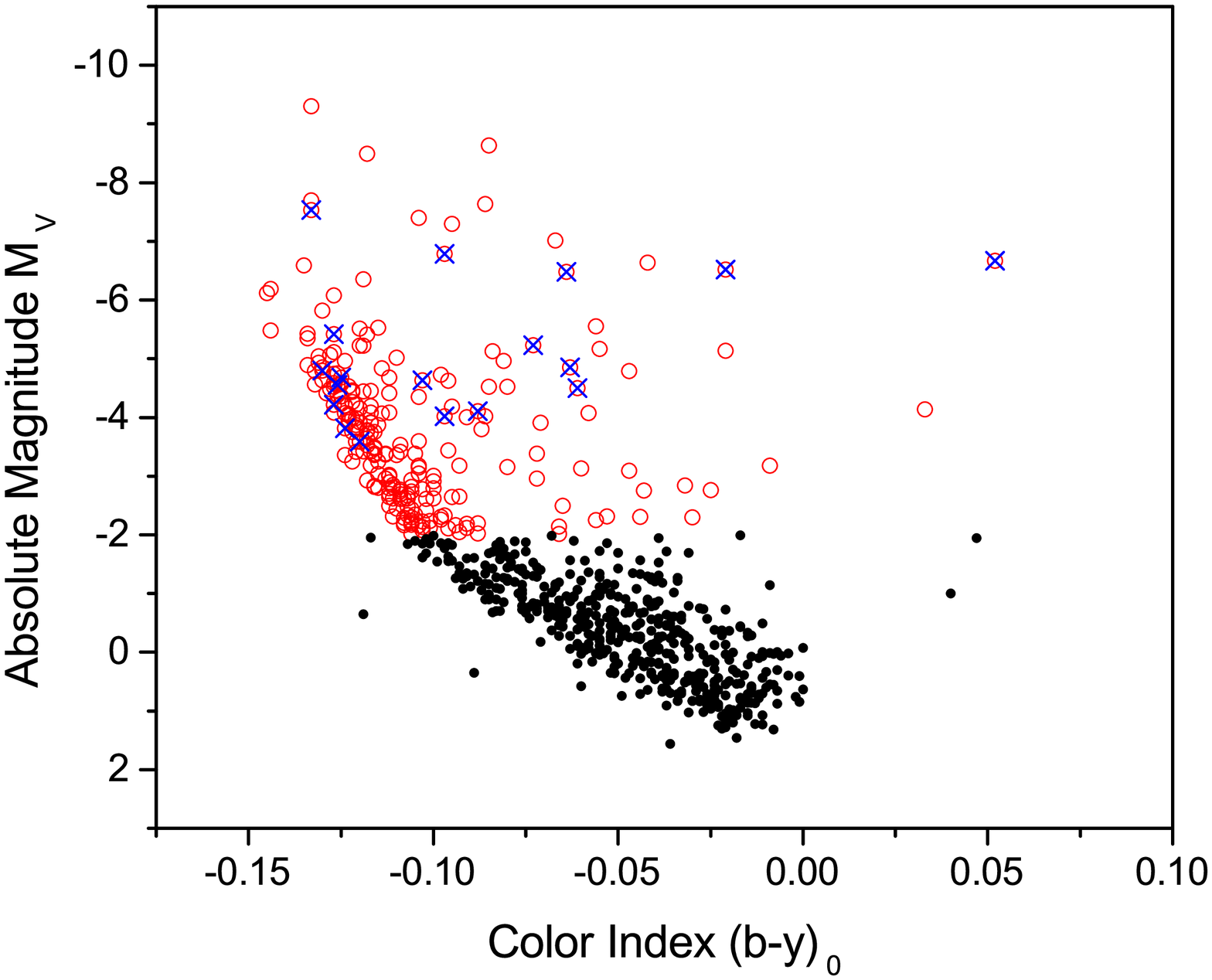}
 \includegraphics[width=18pc]{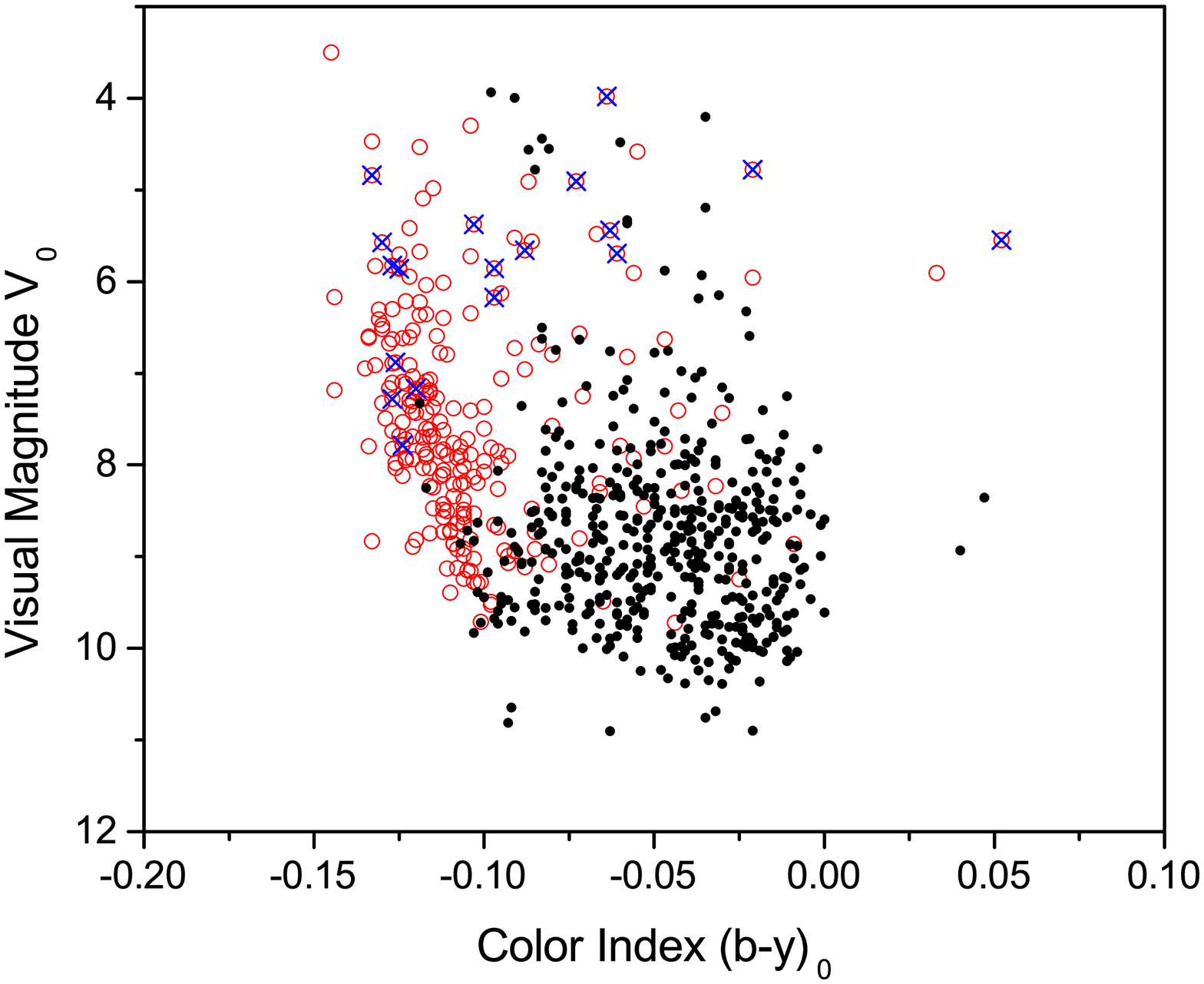}
 \includegraphics[width=18pc]{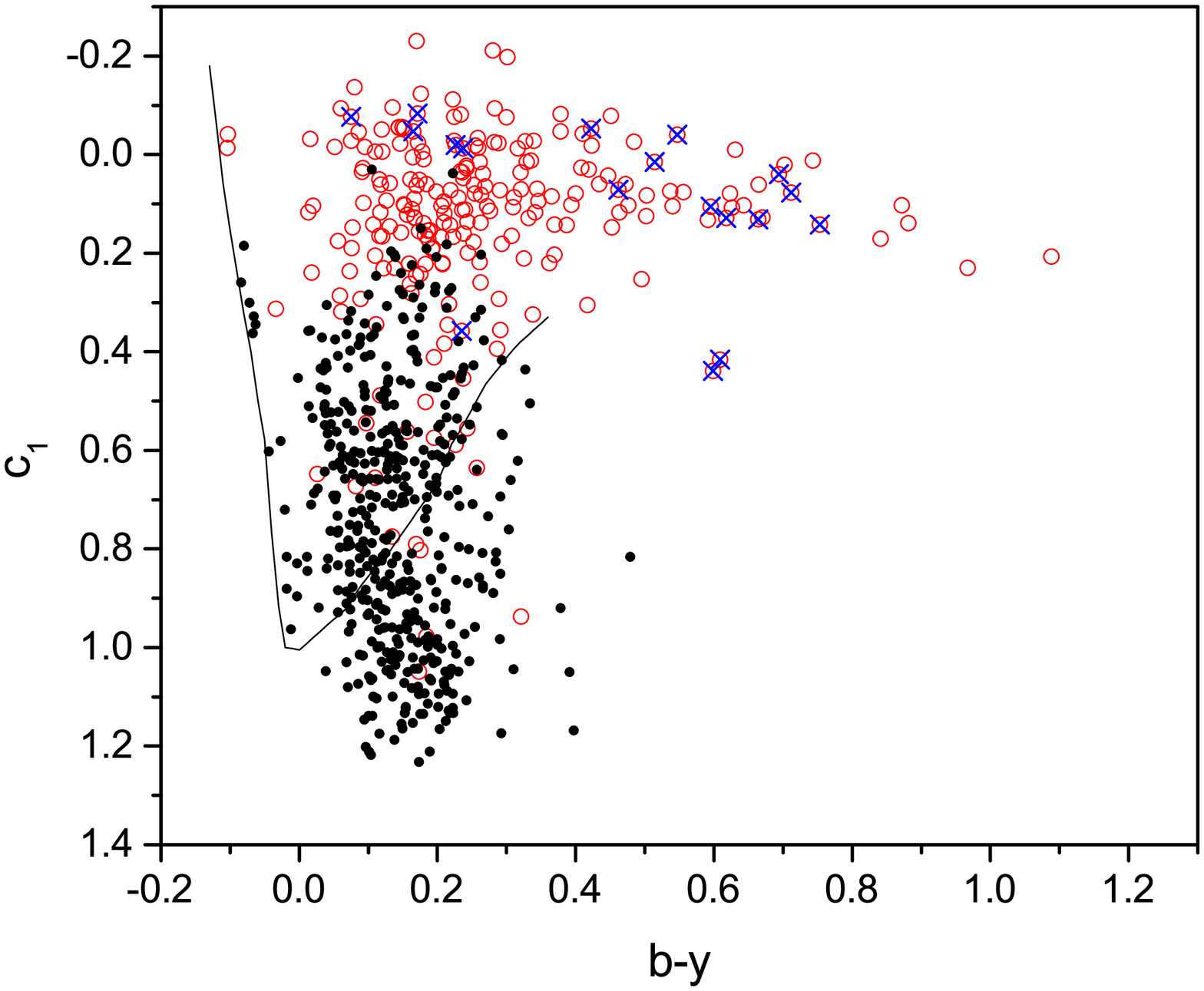}
 \includegraphics[width=18pc]{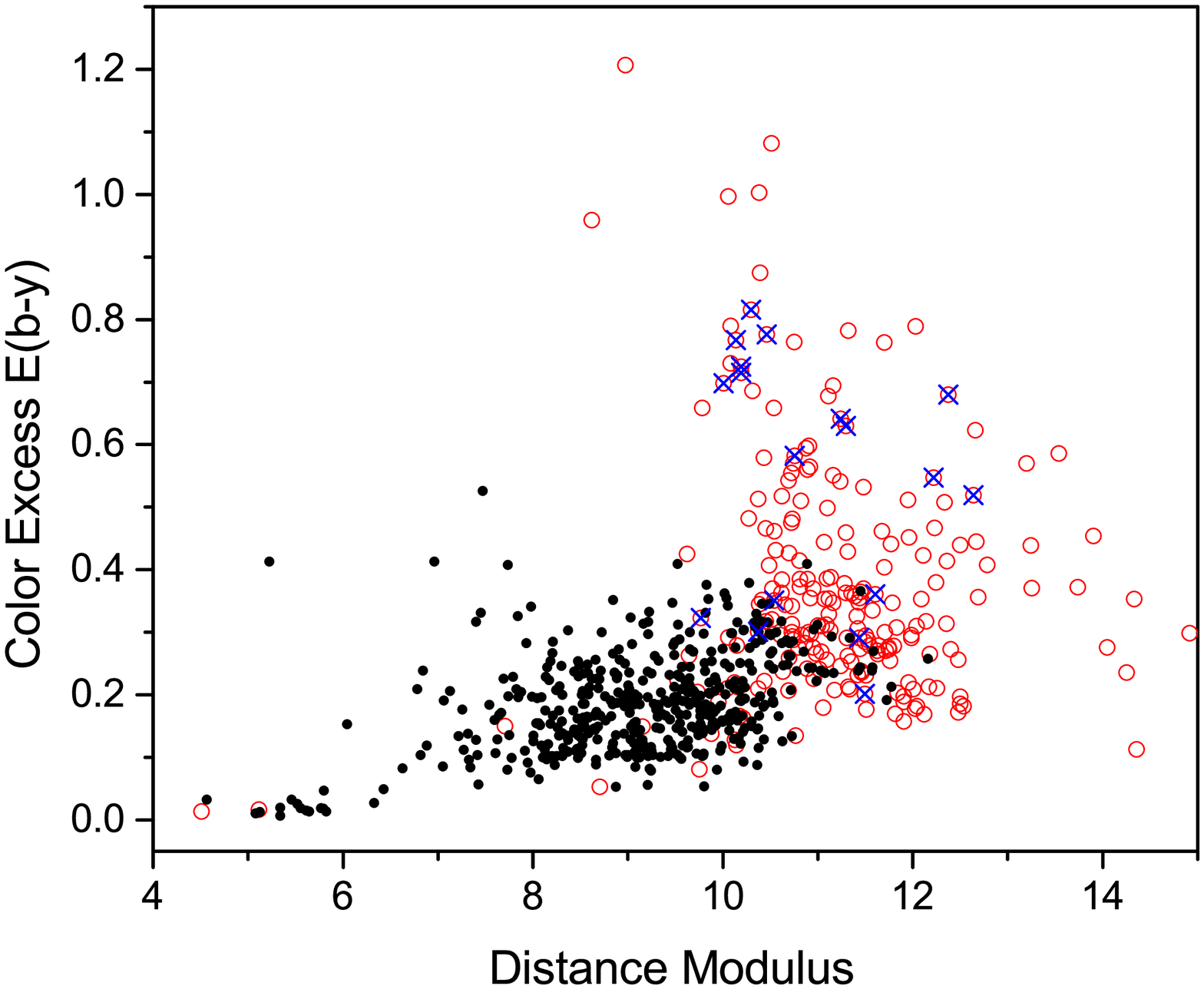}
 \caption[]{Photometric diagrams $M_V$ vs. \by0, $V_0$ vs. \by0 and $c_1$ vs. $b-y$.
   The color excess \Eby\ vs. DM is shown in the last panel.
	 Symbols are the same as in fig.~\ref{fig1}: intrinsically bright stars with $M_V < -2$ mag are
	 marked by red open circles, stars intrinsically fainter than $-2$ mag by black filled circles,
	 stars from the classical Cen OB1 association are marked by blue x-symbols.
	 See the electronic edition for a color version of the figure.}
 \label{fig3}
 \end{figure*}
	
 % Fig 4 *******************************************************
 %
 \begin{figure*}
 \centering
 \includegraphics[width=20pc]{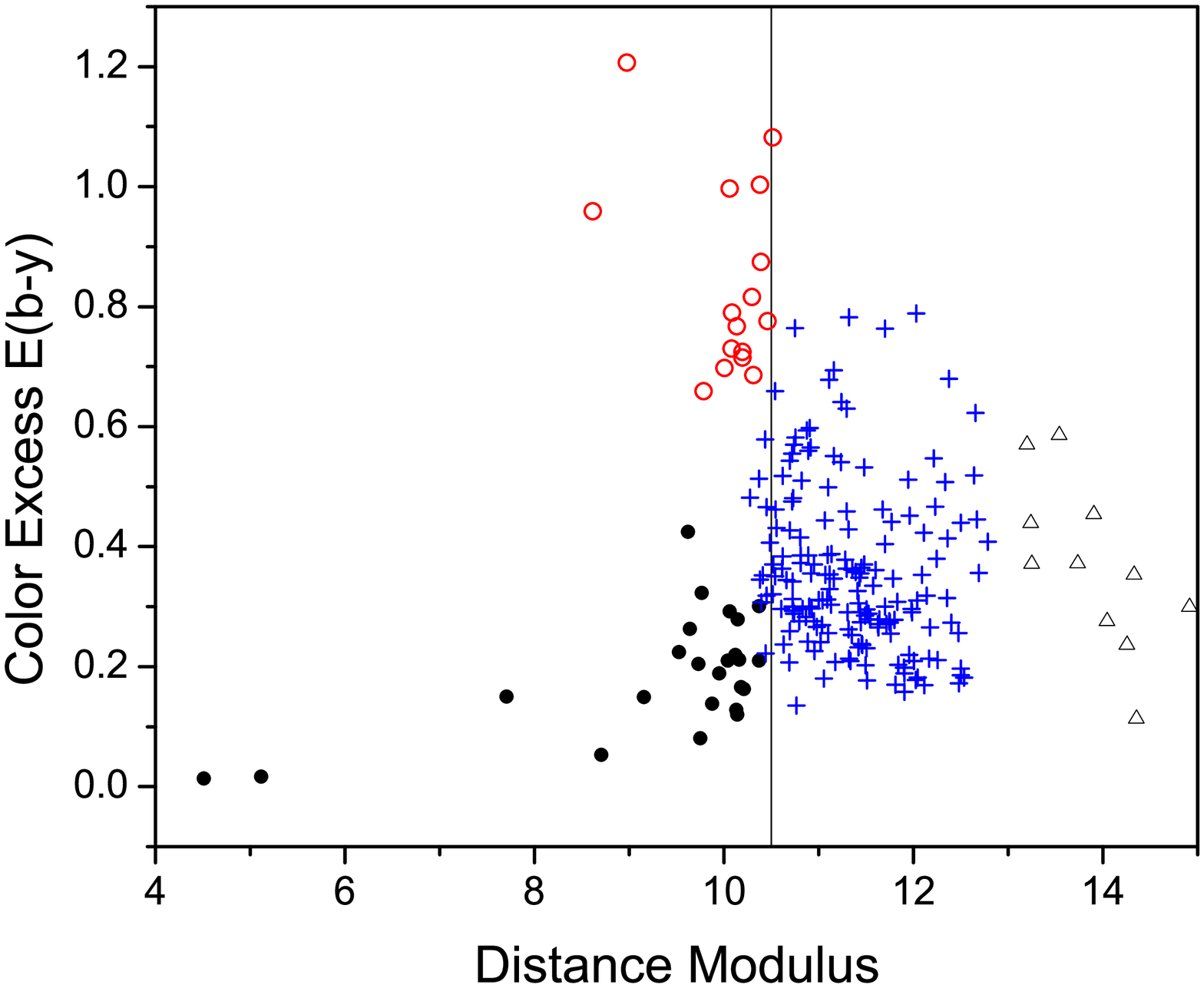}
 \includegraphics[width=20pc]{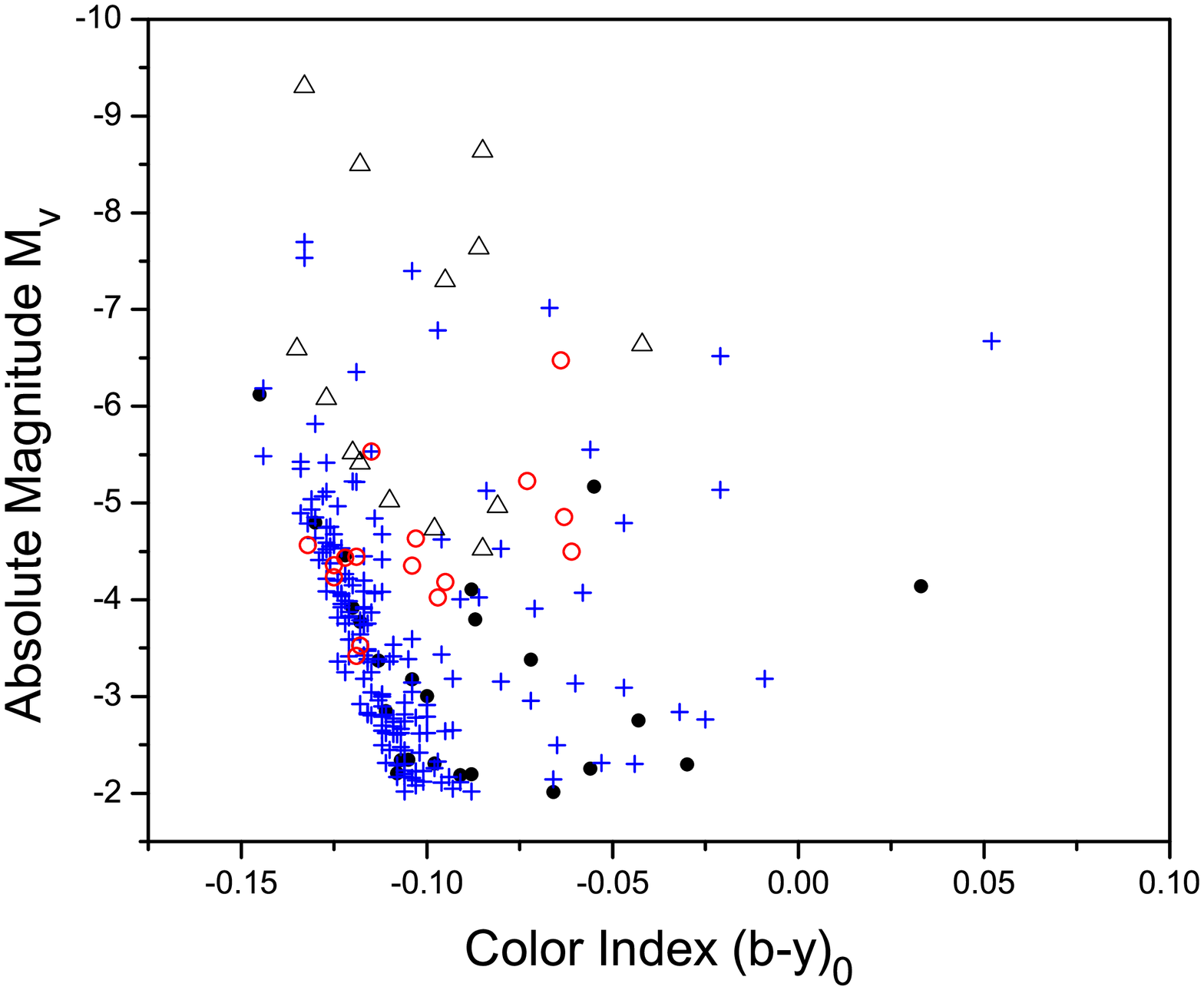}
 \caption[]{$E(b-y)$ vs. DM relation (left) and $M_V$ vs. $(b-y)_0$ for all stars intrinsically
   brighter than \Mv$\ =-2$ mag.
	 Open and filled circles represent stars at a distance around 1 kpc, 
	 blue cross symbols are for stars between 1 and 4 kpc, and triangles are for the most
	 distant stars in the sample.
	 See the electronic edition for a color version of the figure.
 }
 \label{fig4}
 \end{figure*}

 According to their derived distances and photometric quantities, there is a clear separation of intrinsically
 bright, massive, recently-born stars delineating relatively distant sites of star-formation from nearby,
 less massive stars (Fig.~\ref{fig3}).
 A tentative separation of sample stars is made at $M_V=-2$ mag, which reflects well their
 overall properties.
 No clear interarm space is detected toward Centaurus, but rather a continuous
 distribution of stars between nearby Galactic features and the more distant Centaurus
 star-forming complex (see Fig.~\ref{fig3}, Color Excess vs. DM diagram).
 Interstellar extinction towards Centaurus is almost negligible up to 160 pc (DM$\ = 6$ mag).
 At $\sim 1$ kpc a steep increase in extinction is noticeable.
	
 In order to search for coherent groups that can be connected to distant star-forming features we
 removed for the following analysis all stars with $M_V > -2$ mag (marked with black filled symbols on
 Figs.~\ref{fig1} and \ref{fig3}.
	
 Fig.~\ref{fig4} represents only stars with $M_V < -2$ mag., which are relatively distant
 with few exceptions.
 A group of very young stars is noticeable at about 1000 pc, marked with filled and open circles
 depending on their reddening.
 Part of the group (hereafter till the end of the paper marked with open circles), concentrated
 toward average Galactic coordinates $l = 305^\circ, b = 1^\circ$, is quite reddened (\Av $\ = 3$ mag).
 The remaining of the young stars located at 1000 pc (marked with filled circles) are less reddened
 and dispersed across the field.
 Since this is the first time these 40 stars have been distinguished, they are listed in Table~3,
 available only electronically.
 The format of the table follows the format of Table~1.	
	
 The rest of the intrinsically bright massive stars ($M_V < -2$ mag) lie at distances of 1200 to 4000 pc
 (marked in Fig.~\ref{fig4} with plus-symbols), delineating an extended field of recent star-formation.
 Of them, 172  are located between 1200 and 3600 pc (median distance 1820 pc $\pm$583~s.d.; $\pm$44~s.e.).
 Eleven stars marked with open triangles in all plots in Fig.~\ref{fig4} appear to be more distant than DM$\ = 13$ mag.

 We noticed that some of the very massive stars in the sample (\Mv$\ < -5.5$ mag, not shown with separate
 symbols) are detected  towards $l = 306^\circ$ probably tracing a tangential spiral segment in this
 direction.
 Longitude $l = 308^\circ$ defines a tangential direction in the  Galaxy.
 Distant OB stars towards $l = 306^\circ$ have been previously noted by \citet{jac76}.
 \citet{tur85} confirmed the existence of (at least) four distant OB stars (LS 3031, LS 3033, LS 3035
 and LS 3040) background to St 16.
 Complete \uvbyb photometry is available for one of the stars, yielding a distance modulus of 12.6 mag,
 in agreement with that estimated by \citet{tur85}.
		
 The three brightest members of St 16 are included in the \uvbyb sample and yield a median distance modulus
 of $11.45 \pm 0.33$ mag and color excess \Eby\ of $0.38 \pm 0.007$ ($E(B-V) = 0.51$).
 Both estimates agree with previous values based of $UBV$ photometry (see \citealp{tur85};
 V$\acute{\mathrm a}$zquez et al. \citeyear{vaz05}).
 A photometric distance modulus of 11.45 mag (1950 pc) is found for the O7.5~III star HD~115455
 (uvby98 431160111, the brightest cluster member), in agreement with the cluster distance estimates.
 The star is thought to be the exiting source for RCW 75 (see V$\acute{\mathrm a}$zquez et al. \citeyear{vaz05}
 and references therein for details), which provides a reliable distance estimate for this \hii\ region.

 A new distance estimate is made for the cluster NGC 4755 based on 56 cluster members.
 The median and average values for the sample are identical and yield a distance modulus of
 11.48 ($\pm 0.46$~sd; $\pm 0.061$~se).
 The corresponding mean distance is $1968 \pm 65$ pc and agrees closely with the average distance of the
 stars associated with the GSH 305+01$-$24 shell (see Section~\ref{sec:correlation}).
 The average color excess is \Eby\ $ = 0.29\ (\pm 0.03$~sd; $\pm 0.0041$~se).
 These estimates are in perfect agreement with other authors (see \citet{cor13} and the reference therein).
		
 In a summary, the distances derived for nearly 230 massive field stars indicate that the Centaurus
 star-forming field is closer than estimated previously, at a median distance of 1.8 kpc instead of the
 classical 2.5 kpc, the difference likely arising from an overestimation of spectroscopic
 distances in previous studies.
 A portion of stars in the classical Cen OB1 belong to the group detected at 1 kpc, while part is more distant.
 Note that \citet{tur85} provided a similar estimate of $1.88 \pm 0.24$ kpc to Cen OB1 based mainly on
 $H\beta$ photometry of 10 members of Cen OB1 with reliable luminosities.
 The partition of Cen OB1 into five groups by \citet{mel95} is not confirmed, nor are groups of different
 age detected among the stars under consideration.
 The overall impression is that beyond 1 kpc an extended star-forming complex of many OB stars is located.

 % Fig 5 ******************************************************
 %
 \begin{figure}
 \centering
 \includegraphics[width=21pc]{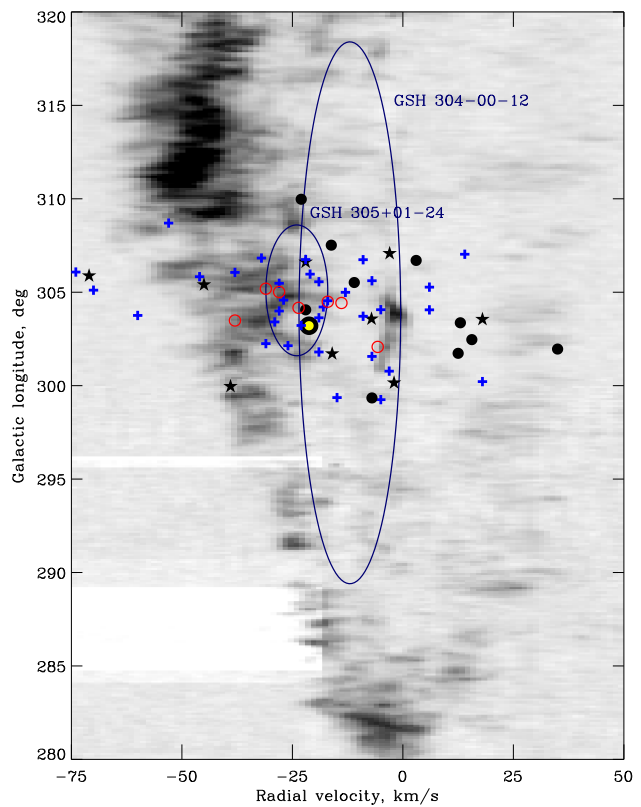}
 \caption[]{Longitude-velocity map, CO J = 1-0 (115 GHz) emission survey of \citet{dam01}.
   GSH 304$-$00$-$12 and GSH 305+01$-$24 are marked by ellipses located at
	 their cataloged positions whose sizes are determined by their angular diameters and velocity widths.
	 As in Fig.~\ref{fig4}, stars of high- and low-reddening at 1 kpc are indicated by the open and filled
	 circles respectively, and blue cross symbols are for stars between 1 and 4 kpc.
	 Intrinsically bright stars with $M_V < -5.5$ mag are shown with black 5-pointed star symbols.
	 The position of NGC 4755 is shown with large yellow circle outlined with black.
	 See the electronic edition for a color version of the figure.
 }
 \label{fig5}
 \end{figure}

 To shed more light on the groups and layers discussed above we collected all available radial velocities
 for these stars. In Fig.~\ref{fig5}, following  \citet{mcc02}, we superimposed them on a
 longitude-velocity ($l,v$) diagram based on $^{12}$CO J = 1-0 (115 GHz) datacube from the CO survey by \citet{dam01}.
 The CO-emission was averaged over the latitude range $|b| < 2$, and the axis range was adjusted to include
 the Carina tangent at $l,v \sim 282^\circ, 0$ km s$^{-1}$ and the Crux-Scutum arm at
 $l,v \sim 310^\circ, -50$ km s$^{-1}$.
 In the figure we show the foreground shell GSH 304$-$00$-$12  at a kinematic distance $1.2 \pm 0.8$ kpc,
 also discovered by \citet{mcc01}.
 As in Fig.~\ref{fig4}, the high- and low-reddened stars at 1 kpc are indicated by open and filled circles,
 respectively, the blue crosses correspond to stars between 1 and 4 kpc, and among them the intrinsically
 brightest stars in the sample ($M_V < -5.5$ mag) are shown by black 5-pointed star symbols.

 As noted by \citet{mcc02}, the GSH 304$-$00$-$12 shell lies in the interarm region and is bounded by
 CO emission from the Coalsack nebula near $l,v \sim 301^\circ, 0$ \mbox{km s$^{-1}$}.
 GSH 305+01$-$24 at $2.2 \pm 0.9$ kpc can be associated with the massive stars around 1.8 kpc, including NGC 4755.
 The highly-reddened group at 1 kpc seems to be located on the near edge of the same shell, while
 the low-reddened layer at 1 kpc seems to be kinematically connected to GSH 304$-$00$-$12.

 %
 %________________________________________________________________

 \subsection{Comparison to other studies}

 As noted earlier, several highly reddened stars from the 1-kpc layer, and also some of the most massive
 and distant stars in the sample are grouped toward $l,b = 305^\circ, 1^\circ$.
 The region toward $l = 305^\circ$ is particularly interesting, containing the G305 \hii\ complex (G305.4+0.1),
 one of the most massive star-forming structures yet identified within the Galaxy \citep[see][]{fai12},
 its central obscured young star clusters Danks 1 and Danks 2, the Wolf-Rayet stars WR 48a,b and $\theta$ Mus,
 other embedded clusters, several \hii\ regions, OH/H$_2$O maser sources, and more (see \citealp{bau09} for a
 detailed description).

 \citet{bau09} present a wide-field photometric study of the Galactic plane towards $l = 305^\circ$,
 revealing a significantly high interstellar absorption (\Av\ $\sim 10$ mag) and an abnormal extinction
 law in the line of sight.
 They found (see also V$\acute{\mathrm a}$zquez et al. \citeyear{vaz05} and \citealp{car09}) three
 different spiral features at increasing distances from the Sun: a spatially spread foreground
 population placed mainly between $350-750$ pc, a second feature situated just behind a very dark
 cloud at $1-3$ kpc and represented by the clusters Danks 1 and 2, and, finally, a third population
 that is traced by embedded clusters located at about $5-7$ kpc.

 These results are in accord with our findings for the field.
 In our sample, the spatially spread foreground population is represented by stars intrinsically
 fainter than $-2$ mag (black filled circles in Fig.~\ref{fig3}).
 The highly reddened stars at 1 kpc are located at the near end of the dust cloud seen
 toward $l = 305^\circ$, at the near edge of the Carina arm.
 The Centaurus complex stretches through the arm, between 1.2 and 3.3 kpc approximately.
 The distant, low-reddened stars we have in our sample lie between 4 and 6.5 kpc, consistent with
 the near end of the Scutum-Crux arm.

 It can be noted that the uncertainties in distance to both embedded clusters and ISM features toward
 the Centaurus-Crux tangent are quite large.
 Recently \citet{dav12} found a kinematic distance to G305 of $4.2 \pm 2.0$ kpc and a spectrophotometric
 distance to Danks 1 and 2 of $3.48^{+0.91}_{-0.71}$ kpc $2.61^{+0.82}_{-0.61}$ kpc, respectively.
 By comparison, the \uvbyb distances delineate at least the near side of the Centaurus star-forming
 field more precisely.

 % Fig 6 *******************************************************
 %
 \begin{figure*}
 \centering
 \includegraphics[width=20pc]{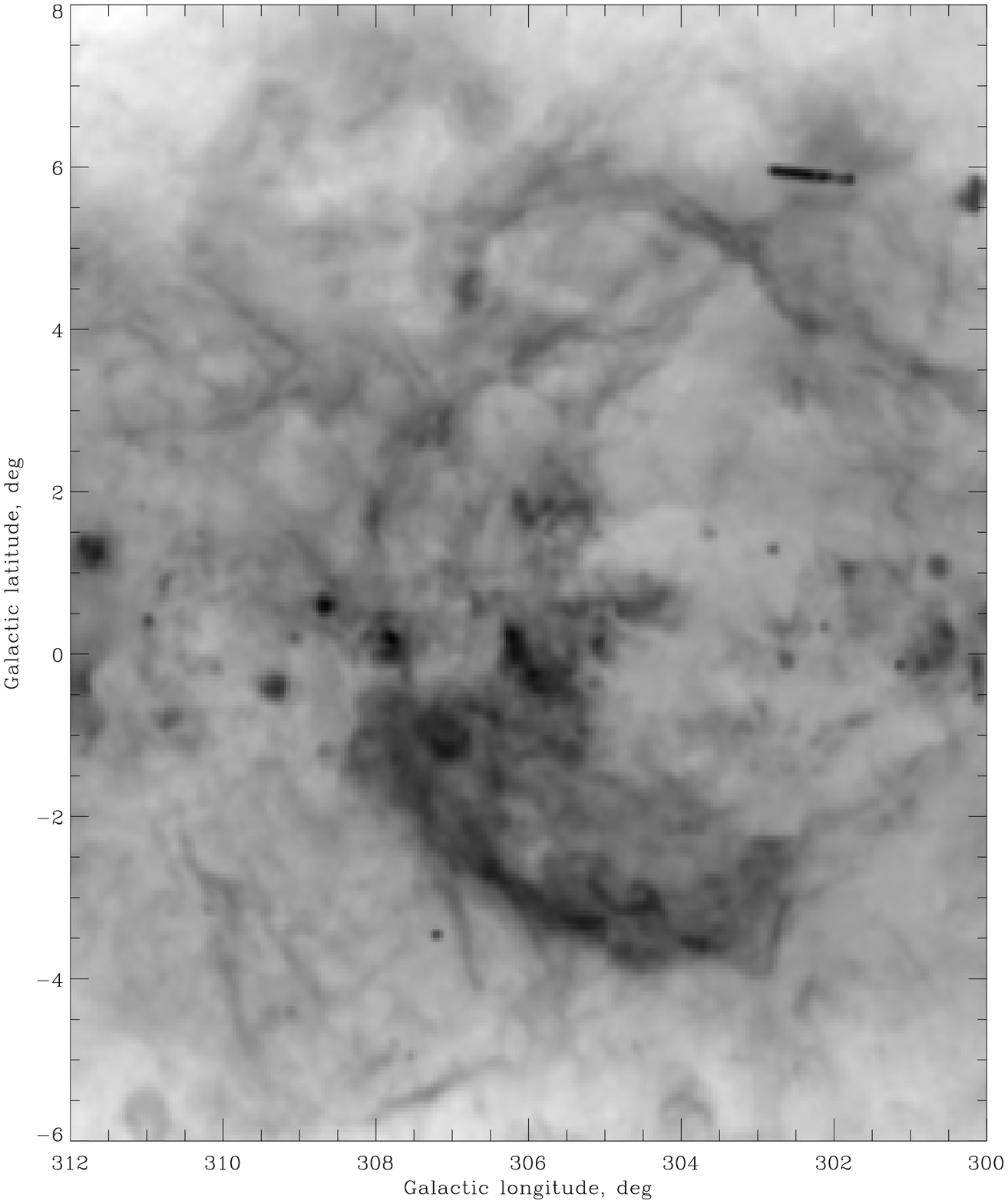}
 \includegraphics[width=20pc]{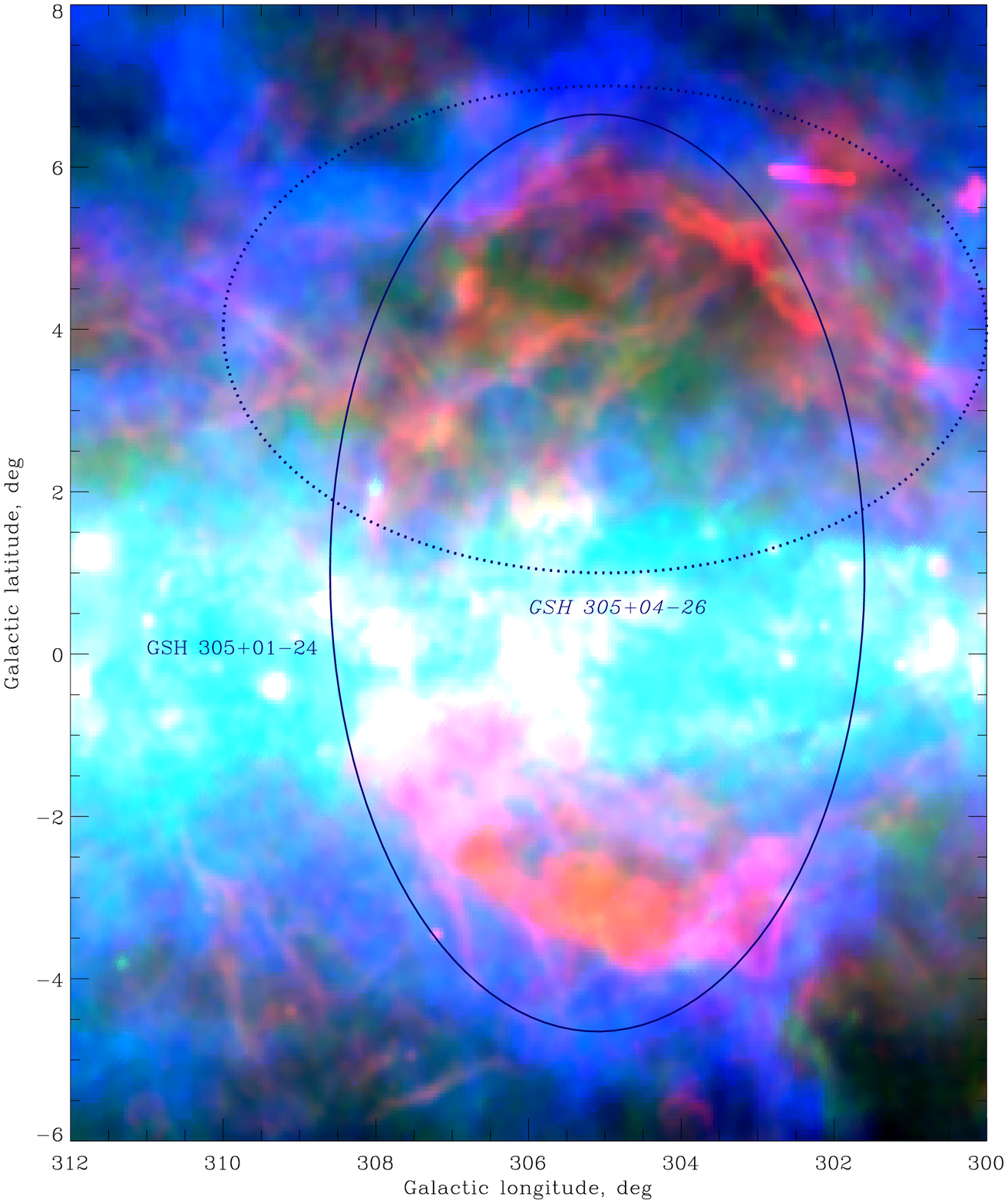}
 \caption[]{Left: H$\alpha$ emission (data by \citealt{fin03}) toward the Coalsack Loop.
	 Right: H$\alpha$ emission (red) combined with \hi\ 21-cm emission (blue) and 100 $\mu$m DIRBE
	 far-IR emission (green) for the same region.
   The solid-line ellipse outlines the \hi\ shell GSH 305+01$-$24 discovered by \citet{mcc01}.
	 The \hi\ 3D-datacube of the region were taken from {\em The Southern Galactic Plane Survey} \citep{mcc05}.
   Here the $-24$ km s$^{-1}$ channel of \hi\ emission is shown.
	 The position of the shell GSH 305+04$-$26 (labeled in italics) recently proposed by \citet{cor12}
	 is outlined by a dotted line.
	 See text for discussion.
	 See the electronic edition for a color version of the figure.
 }
 \label{fig6}
 \end{figure*}		

 %	
 %________________________________________________________________	
	
 \section{The census of OB stars within GSH 305+01-24}\label{sec:correlation}

 As already noted, the  massive, intrinsically luminous stars found between 1 and 4 kpc delineate an
 extended star-forming field.
 Of this sample, 133 stars are projected over the sky area covered by GSH 305+01$-$24 and are located at
 an average distance of 1.8$\pm$0.4 kpc (mean DM $=11.3\pm 0.60$, median DM $= 11.24$).
 We consider this "shell subsample" as part of the massive stellar population of GSH 305+01$-$24
 that matters from the wind injection energy viewpoint.

 In order to estimate the completeness of this shell subsample, first we compared the  selection to
 the $Hipparcos$ catalog. Only 5 stars were found there, all of them with very small or negative parallaxes,
 as it should be expected for stars more distant than 1 kpc.
 This confirms our statement that the shell subsample is not contaminated by any nearby young stars.
 This is especially important for the Centaurus star-forming field, which is background to the nearby
 Sco-Cen association.

 Nearly 70\% of the shell subsample stars have HD numbers as primary designations.
 Because of some spectral type discrepancies in the B8-A2 range that appear the HD catalog, it is difficult
 to estimate the completeness precisely.
 For example, HD 112537 is classified as B9 in the HD catalog, but as A1V in SIMBAD and thus omitted in
 our sample which is focused on stars of spectral types earlier than A0.
 An estimate reflecting these discrepancies indicates that the shell subsample is $\sim 90$\% complete
 to the magnitude limit of the HD catalog.
 Similar estimate was obtained for the completeness of stars having only BD/CD/CPD identifiers.

 These estimates point out  that the completeness of the sample within the shell is about 85-90\% to a
 limiting magnitude of $11.5-12$ mag for spectral types O-B9.
 
 %	
 %________________________________________________________________

 \subsection{Observational properties of GSH 305+01$-$24 shell}

 GSH 305+01$-$24 was first recognized as a large ring of H$\alpha$ emitting nebulosity of
 about $10^\circ$ in diameter, surrounding the Coalsack Nebula and was referred to as
 "The Coalsack Loop" \citep[][see our Fig.~\ref{fig6} left]{wal98}.
 Based on observations with the Parkes Radio Telescope it was then identified with a large \hi\ supershell
 in this direction \citep{mcc01}.
 The parameters of GSH 305+01$-$24 are summarized by \citet{mcc02} as follows:
 $l,b = 305^{\circ}\!\!.1, +1^{\circ}\!\!.0$; $\Delta l \times \Delta b = 7^\circ \times 11^{\circ}\!\!.3$;
 central radial velocity $v_\mathrm{LSR} = -24$ km s$^{-1}$; velocity full width  $\Delta v  = 14$ km s$^{-1}$,
 and expansion velocity $v_\mathrm{exp} \sim 7$ km s$^{-1}$.
 The shell extends between $-5^\circ$ and $+7^\circ$ across the Galactic plane.
 According to \citet{mcc01}, $v_\mathrm{LSR} = -24$ km s$^{-1}$ implies a kinematic distance of
 $2.2\pm0.9$ kpc and  physical dimensions of $280 \times 440$ pc.
 Since the estimated distance of GSH 305+01$-$24 at 2.2 kpc is consistent with that of Cen OB1 at 2.5 kpc,
 as determined by \citep{hum78}, the \hi\ shell has been identified as blown out by the classical Cen OB1
 association.
 However, the kinematic distance of the shell is equally consistent with the revised distance to Cen OB1.

 The distance to the shell subsample obtained here  places the shell at a photometric distance of $1.8\pm0.4$ kpc.
 Within the errors, this estimates agrees  with the kinematic distance $2.2\pm0.9$ kpc and implies shell size
 $a \times b = 229 \times 360$ pc.
 The harmonic equivalent radius is calculated as $R_s = 0.5\sqrt{a \times b} \approx 143.5 \pm 32$ pc.

 Fig.~\ref{fig6} (right) presents the \hi\ radio-data (obtained from the archive of
 {\em The Southern Galactic Plane Survey} (\citealt{mcc05}) combined with the H$\alpha$ data
 by \citet{fin03} and 100 $\mu$m DIRBE far-IR data \citep{sfd98}.
 The GSH 305+01$-$24 is outlined with a solid ellipse. The Coalsack Loop shows a striking similarity
 between the \Ha\ (inside) and \hi\ (outside) emission morphologies.
 The dotted ellipse in Fig.~\ref{fig6} (right) represents the GSH 305+04$-$26 shell, recently
 suggested by \citet{cor12}.
 Based on \hi\ data, the latter authors argued that the two halves of GSH 305+01$-$24, located above
 and below the Galactic plane (their features A and B), are unrelated \hi\ structures and do not
 manifest different parts of the same feature, as suggested by \citet{mcc01}.
 The H$\alpha$ image, however, does not support this conclusion.
 The ISM emissions at different wavelengths and the distribution of the stellar content of the shell
 confirm the existence of only one shell, with \hi\ outward, H$\alpha$ inward, and a cavity full
 of dust IR emission.
 The finding of \citet{cor12} of two unrelated \hi\ shells above and below the Galactic plane are
 not confirmed by the present results.

 %	
 %________________________________________________________________

 \subsection{The wind-blown energy deposit of the most massive stars within GSH 305+01$-$24}

 Ionized shells around OB associations are usually connected to the initial stages of star cluster
 blown bubbles \citep[e.g.][]{sil08,sil13} and are suggested to be progenitors of the larger
 \hi\ supershells \citep[see][]{rel07,sil13}.

 The mechanical luminosities of the stellar winds of massive MS stars influence the size and  shape
 of the \hi\ superbubbles.
 The final size of a superbubble created in a molecular environment during the massive stars MS stage
 depends on the stellar mass loss, wind velocity, and environmental density.
 Since stars in the mass range 8 to $25-30\ M_\odot$ will end their lives in a RSG phase,
 %and will not launch a further WR wind
 the extent of molecular gas cavity is largely determined by the merged wind bubbles
 (i.e. combined bubble blown from the contribution of the output of all massive stars).

 The energetics of the \hi\ shells are usually characterized by the expansion energy, $E_\mathrm{exp}$,
 defined as the equivalent energy instantaneously deposited at the shell center to account for the shell’s radius,
 $R_s$, and  also by the current rate of expansion, $v_\mathrm{exp}$.
 As in \citet{mcc01}, to estimate the expansion energy of GSH 305+01$-$24 we have used the expression \citep{che74,hei79}:
 \[
 E_\mathrm{exp} = 5.3 \times 10^{+43} n_0^{1.12} R_s^{3.12} v_\mathrm{exp}^{1.4} \approx (3.5 \pm 1.9) \times 10^{51}\,
 \mathrm{ergs,}
 \]

 \noindent where the ambient density $n_0$ is adopted to be 1 cm$^{-3}$, a typical value (see e.g. \citealt{loz99}
 for details).

 In absence of an observed power source, the real age of the shell could not be determined.
 Instead, an age estimate based on dynamic properties, or the dynamic age, can be quoted.
 Following \citet{cor12} who used the \citet{wea77} standard model for a thin expanding shell created by a continuous
 injection of mechanical energy, we derive the dynamic age $t_\mathrm{dyn}$ of GSH 305+01$-$24 as follows:
 \[
 t_\mathrm{dyn} = 0.55 \times 10^6 R_s/v_\mathrm{exp} \approx 13.15 \pm 2.95\ \mathrm{Myr.}
 \]

 % Fig 7 *******************************************************
 %
 \begin{figure}
 \centering
 \includegraphics[width=20pc]{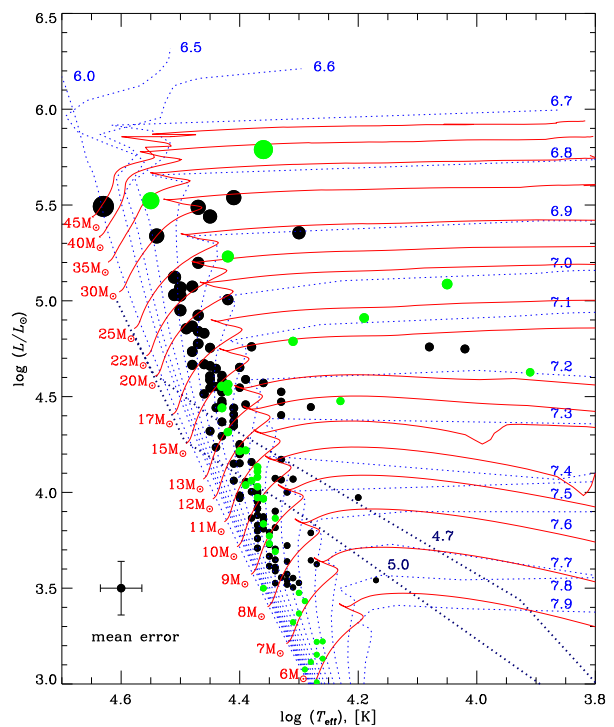}
 \caption[]{The HR diagram for the stellar content of GSH 05+01$-$24.
	 Stars of different masses are represented with filled circles of different diameters.
	 Black circles are used for the shell subsample, and green circles denotes NGC 4755 massive members.
	 See the text for details.
	 See the electronic edition for a color version of the figure.
 }
 \label{fig7}
 \end{figure}
	
 % Fig 8 *******************************************************
 %
 \begin{figure}
 \centering
 \includegraphics[width=20pc]{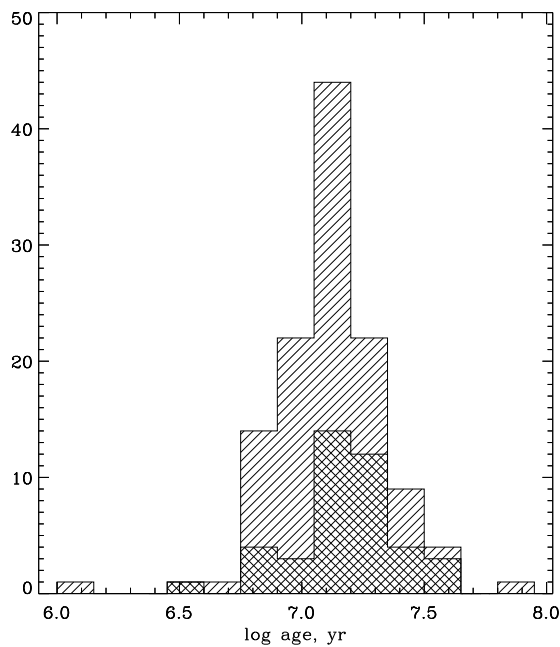}
 \caption[]{Histogram of stellar ages for the shell subsample (119 stars, diagonal pattern)
	 and NGC 4755 sample (41 stars, cross-lined pattern).
 }
 \label{fig8}
 \end{figure}

 Next we calculate whether the stars within the shell with masses $M \gtrsim 8 M_\odot$ are able to
 contribute a sufficient wind injection energy in order to explain the observed size and expansion velocity.

 Fig.~\ref{fig7} presents the HR diagram for the shell subsample with masses $M \gtrsim 8 M_\odot$.
 NGC 4755 is also included.
 The effective temperatures $T_\mathrm{eff}$ and bolometric corrections are obtained via calibrations in
 terms of $c_0$ and [u-b] \citep[see][]{dav77, cod76}.
 Evolutionary tracks (mass range $6 \div 45 M_\odot$) and isochrones (log age range $6.0 \div 7.9$) by \citet{eks12}
 extracted via the {\em Geneva stellar models:
 interactive tools}\footnote{http://obswww.unige.ch/Recherche/evoldb/index/} are overplotted.
 We use models with initial metallicity $Z_\mathrm{init} = 0.014$ (solar) and initial rotation rate
 $v/v_\mathrm{crit} = 0.40$.
 Two isochrones for pre-main sequence stage (log age 4.7 and 5.0, \citet{cla12}) are also shown with thick dotted lines.
 The individual stellar ages are extracted from the models.
 The individual log age distributions for the shell subsample and NGC 4755 sample are shown in Fig.~\ref{fig8}.
 We estimate an average age of the shell subsample of $7.087 \pm 0.177$ or $\approx 12.2 \pm 4.09$ Myr,
 in good agreement with the dynamic age $t_\mathrm{dyn}$ of the shell ($13.15 \pm 2.95$ Myr), calculated above.
 For NGC 4755 the average log age derived by us is $7.183 \pm 0.230$ or 15.2 Myr.

 The estimates for the distance and age of NGC 4755 we derived are in a very good agreement with previous results.
 \citet{cor13} give distance to the cluster $1.9 \pm 0.5$ kpc, and age $\sim 15$ Myr.
 \citet{bon06} estimated cluster's age of $14 \pm 2$ Myr and concluded that the star formation in
 NGC 4755 began $\approx 14$ Myr ago and proceeded for about the same length of time.
 \citet{lyr06} found a value of $2040 \pm 250$ pc for the distance to the cluster and $7.1 \pm 0.2$
 for the log age.
 WEBDA database\footnote{http://www.univie.ac.at/webda/Welcome.html} lists log age 7.216 or 16.4 Myr and distance 1976 kpc.

 One of the main results of this work is that NGC 4755 is practically at the same distance as the shell subsample.
 The mean radial velocity of NGC 4755 $-21.24 \pm 0.89$ km s$^{-1}$ \citep{mer08} coincides well with the
 mean radial velocity of GSH 305+01$-$24.
 These are firm arguments that NGC 4755 and the shell subsample form the main part of the massive star population
 with in the shell that should be taken into account when studying the energetics and star-formation
 history of this region.

\begin{figure*}
 \centering
 \includegraphics[width=20pc]{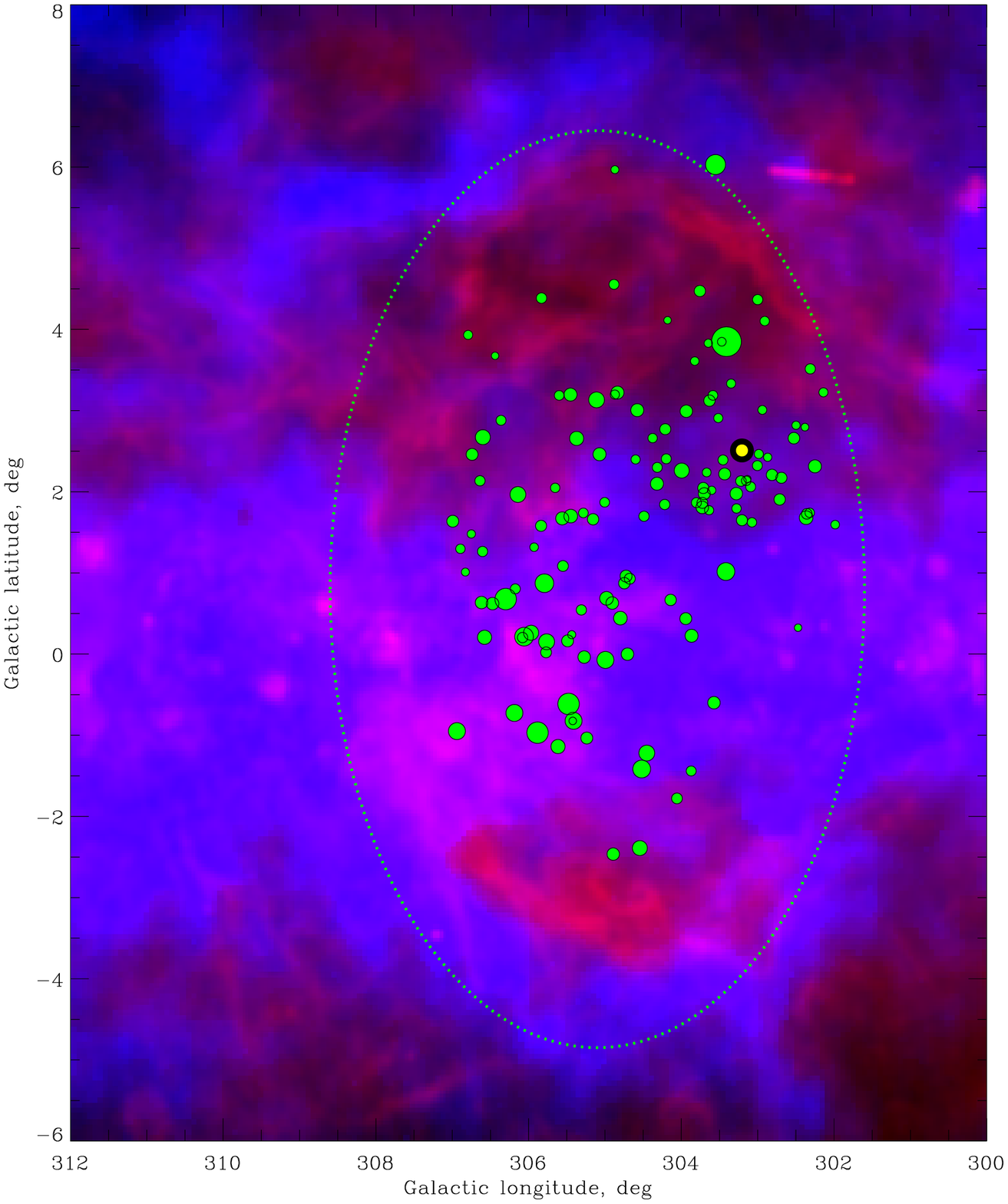}
 \includegraphics[width=20pc]{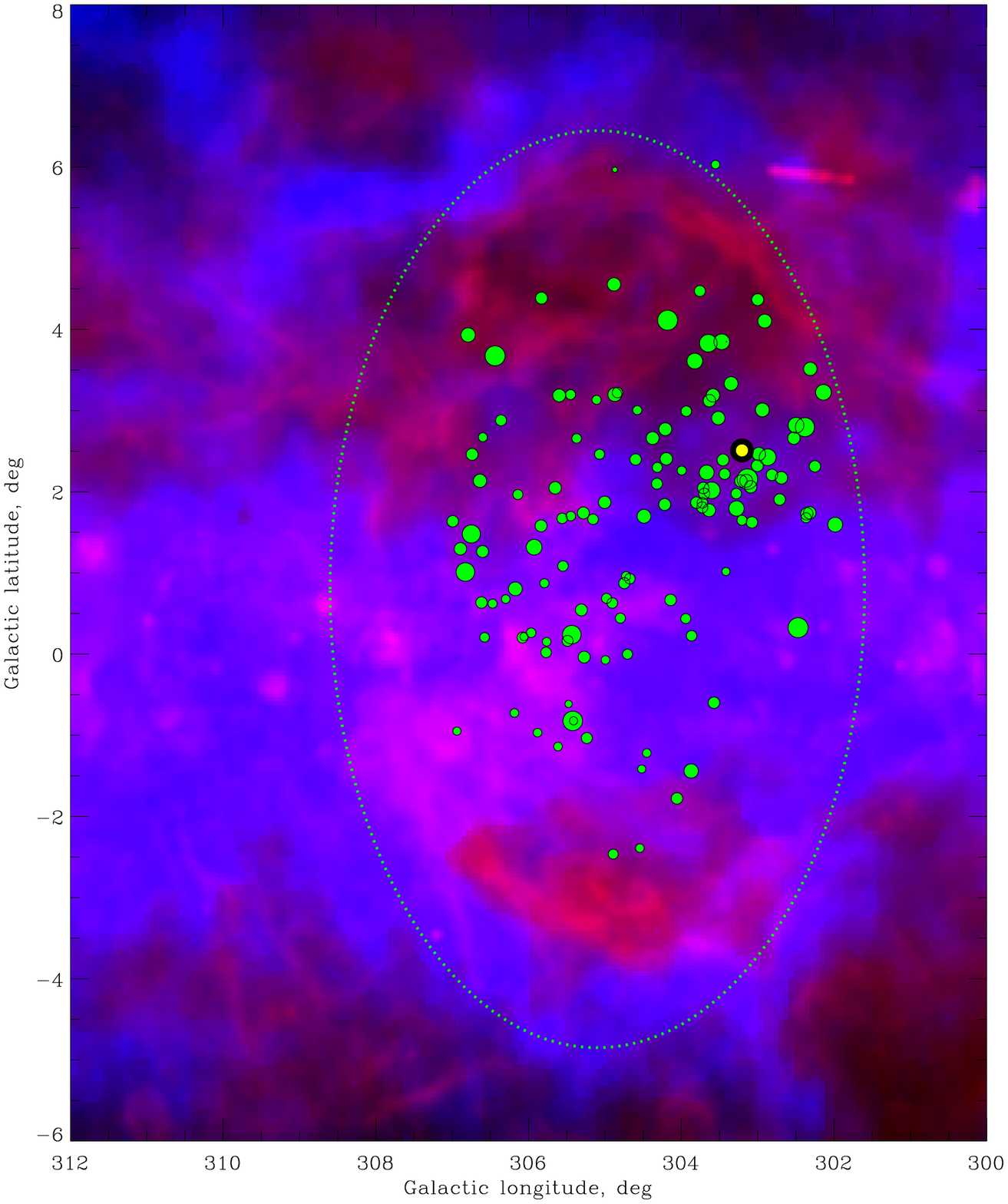}
 \caption[]{
   Massive stars within the Coalsack Loop.
	 H$\alpha$ emission is shown in red and \hi\ 21-cm emission ($-24$ km s$^{-1}$ channel) is in blue.
	 Left: Symbol size is proportional to the stellar mass.
	 Right: Symbol size is proportional to the stellar age.
	 The position of NGC 5755 is shown with yellow circle outlined with black.
	 See the electronic edition for a color version of the figure.
 }
 \label{fig9}
 \end{figure*}

 In Fig.~\ref{fig9} all stars included in the HR diagram (and thus with masses and ages calculated)
 are plotted against the Coalsack Loop.
 As in Fig.~\ref{fig6}, the H$\alpha$ emission (red) is combined with \hi\ 21-cm emission
 (blue, $-24$ km s$^{-1}$ channel).
 The doted-line  ellipse outlines the \hi\ shell GSH 305+01$-$24.
 On the left panel the stars are represented with symbols proportional to their mass as estimated from
 the Geneva stellar models.
 On the right panel the symbol size is proportional to star's estimated age.
 One can follow an age gradient over the Coalsack Loop with NGC 4755 being the oldest.
 A continuous star-formation might be taking place within the shell with the youngest, most massive stars
 located at the periphery of shell.
 The stars  appear to trace the high H$\alpha$ emission, avoiding it at the same time.
 The projected coincidence of the OB stars with the shell and the similarities between the shell's morphology
 and the OB-star distribution indicates a strong interaction of the stellar winds with the superbubble
 material, as also suggested by \citet{mcc01}.

 The kinetic energy deposit into the surrounding media can be estimated based on stellar wind luminosities
 and stellar ages.
 Only stars with masses $M \gtrsim 8 M_\odot$ were taken into account (113 stars from the shell subsample
 and 30 member stars of NGC 4755).
 The parameters involved in the calculations are the wind mass-loss rate, $\dot{M}$, which is the amount of mass
 lost by the star per unit time, and the terminal (asymptotic) velocity $v_\mathrm{inf}$ or the velocity of
 the stellar wind at a large distance from the star.
 The log values of $\dot{M}$ in units [$M_\odot$ yr$^{-1}$] were obtained  from the models.
 The wind terminal velocity can be estimated
 from \citet[eq. 2]{lei92} as a function of luminosity, mass, and $T_\mathrm{eff}$ of the star
 (solar metallicity was adopted).
 With these parameters in hand, the wind energy production of each star during its MS lifetime, $E$, was estimated as
 \[
 E = \frac{1}{2} \dot{M} v_\mathrm{inf}^2 \tau_\mathrm{ms}
 \]

 \noindent (see eq. 1 in \citealp{che99}) where $\tau_\mathrm{ms}$ is the main-sequence age of the star,
 obtained from the models.
 Thus the integrated wind energy deposit of all stars was found to be $E_\mathrm{wind} \approx 2.95 \times 10^{51}$ ergs,
 in quite good agreement with shell's expansion energy $E_\mathrm{exp} \approx (3.5 \pm 1.9) \times 10^{51}$ ergs
 calculated above.

 Ionized shells may present the inner skins of neutral shells if the number of ionizing photons emerging
 from the central cluster is not sufficient in order to ionize the whole shell.
 It looks like the observed shell passes by this stage at present time.

 % Fig 9 *******************************************************
 %
	
 Using a basic model in which energy is supplied first by stellar winds and then by supernovae, \citet{rel07}
 extrapolated the properties of these shells in time, finding results in good overall agreement with the
 observed properties of the \hi\ shells after times of a few $10^7$ yr.
 They also discussed that, if the formation of the OB stars occurs on shorter time scales, the stars will no
 longer be located at the centers of the \hi\ shells \citep[see also][observations reported for the SMC]{hat05}.
 Our results appear to support that.
 Although the massive stellar content of GSH 305+01$-$24 appears shifted towards the near edge of the shell,
 the large uncertainties in the shell kinematic distances do not allow firm conclusions.

 %	
 %________________________________________________________________

 \section{Concluding remarks}\label{sec:concl}

 H$\alpha$, \hi , CO, and 100 $\mu$m emission in the direction of the Centaurus star-forming field is used
 to explore the morphology of the Galactic supershell GSH 305+01$-$24.
 A superposition of large-scale features of the disk of the Milky Way and the supershell GSH 305+01$-$24
 which extends perpendicularly to the disk is clearly outlined.
 The morphology of H$\alpha$ and \hi\ features are consistent with
 the location of GSH 305+01$-$24 as delineated by \citet{mcc01}.
 The positional coincidence of OB stars and the shell along with similarities between the shell's
 morphology and the distribution of OB stars indicates a strong interaction of the
 stellar winds and the bubble material.
 The precise photometric distances obtained here revealed the presence of previously unrecognized groups of stars,
 and led to a complete revision of the OB-star distribution in the Centaurus star-forming complex.
 The young  stellar component beyond 1 kpc occupies the cavity of an extended \hii\ region that is surrounded
 by the \hi\ GSH 305+01$-$24 supershell.
 We identified this component at about $85-90$ \% completeness up to $11.5-12$ mag.
 We show that these stars  are able to contribute a sufficient wind injection energy in order to explain
 the observed size and expansion velocity.
 A previously undetected layer of very young stars was also identified 1 kpc and its connection to both
 GSH 305+01$-$24 and 304$-$00$-$12 \hi\ shells was investigated.

 %
 %________________________________________________________________
	
 \begin{acknowledgements}

 This work is supported by the National Science Foundation grant AST-0708950.
 N.K. acknowledges support from the SNC Endowed Professorship at the
 University of Wisconsin Oshkosh.
 K.M. acknowledges support from the Wisconsin
 Space Grant Consortium.
 V.G. acknowledges support by the Bulgarian National Science Research Fund under the grant DO 02-85/2008.
 This study made use of the NASA Astrophysics Data System, SIMBAD database, {\em Centre de
 Donn$\acute{\mathrm e}$es Stellaires}\footnote{http://cdsweb.u-strasbg.fr/},
 WEBDA open cluster database operated at the Institute for Astronomy of the University of Vienna, and the
 NASA's {\em SkyView} facility\footnote{http://skyview.gsfc.nasa.gov/} located at NASA
 Goddard Space Flight Center \citep[see][]{mcg98}.
 We acknowledge the use of the Southern H$\alpha$ Sky Survey Atlas (SHASSA), which is supported by the
 National Science Foundation \citep{gau01}.
 We acknowledge the use of The Southern Galactic Plane Survey \citep[see][]{mcc05}.
 We are grateful to Sergiy Silich and David Turner for numerous comments that greatly improved the manuscript.
 We are thankful to the referee, Dr F. Comer{$\acute{\mathrm o}$}n, for many valuable comments that greatly
 improved the paper.

 \end{acknowledgements}
	
 %
 %______________________________________________________________

 \bibliographystyle{aa}
 \bibliography{AA2013_21454}
	
 \end{document}